\documentclass[runningheads]{llncs}
\newcommand{\Lstar}{$\mathsf{FOTL_1}$}

\newcommand\beq{\begin{equation}}
        \newcommand\enq{\end{equation}}

\newcommand{\RightBox}{{\phantom{a}}\hfill $\Box$ \\}

\newcommand{\diamin}{\Diamond\kern-0.5em{\raisebox{.25ex}{\rm -}}\kern0.175em}

\newcommand{\To}{\Rightarrow}

\newcommand{\ldot}{{\rm <}\kern-0.37em{\raisebox{.25ex}{\bf .}}\kern0.375em}

\usepackage{packages}
\begin{document}

\title{Verification of Unbounded Client-Server Systems with Distinguishable Clients}
\author{author}
\author{Ramchandra Phawade \inst{1} \and S Sheerazuddin \inst{2} \and Tephilla Prince \inst{3}}

\authorrunning{{Phawade} et al.}

\institute{ IIT Dharwad, India \email{prb@iitdh.ac.in} \and
    NIT Calicut, India \email{sheeraz@nitc.ac.in} \and
    IIT Dharwad, India \email{tephilla.prince.18@iitdh.ac.in
    }
}

\maketitle
\begin{abstract}
    Client-server systems are a computing paradigm in concurrent and
distributed systems. We deal with  unbounded client-server systems (UCS) where all clients are of the same type, interact with a single server and they may enter and exit the system dynamically. At any point in time, the number of clients is bounded, but their exact number is unknown and dynamic.
To model these systems, simple Petri nets are not directly usable,
so we use unbounded $\nu$-nets~\cite{VelardoF07}. Owing to the distinguishability of clients in UCS, it is not straightforward
to express their properties in LTL or CTL.
To address this, we propose the logic {\Lstar}, a monodic fragment of Monadic First Order Temporal Logic (MFOTL)~\cite{HWZ01}. In this work, we provide the SMT encodings of $\nu$-nets and {\Lstar} to do Bounded Model Checking (BMC).
We also build an accompanying open source tool to perform BMC of UCS.
\end{abstract}

\section{Introduction}\label{sec:intro}
Client-server systems are a computing paradigm where work loads are distributed by the service providers called server, to the service requesters called clients, or alternatively, the clients request resources from a server. There are variations to this model, where there are single servers and multiple clients, multiple servers and clients, and communication among the various entities, passive servers and communicating clients~\cite{ChristophidesHKKTX01}. For instance, stock markets and cryptocurrency exchanges with an unbounded number of investors, multiplayer games where the players are not known apriori can be viewed as client-server systems with unbounded agents which we refer to as unbounded client-server systems (UCS). We focus on UCS where clients are distinguishable and there is concurrency in their interactions with the single server.

The larger goal is to formally verify properties of UCS.
Petri nets~\cite{murata89}, a well-studied model of concurrency are not suitable since they do not allow handling of distinguishable clients in a straightforward manner. Hence, the first challenge is to find a suitable formal model for UCS. To express the properties of these systems, traditional logics such as LTL or CTL~\cite{P77} are not natural. Therefore, the second challenge is to find suitable logics for expressing properties of UCS where the number of clients is not known a priori and the clients are distinguishable. In order to verify properties of UCS, we employ model checking, which can be used in the design stage of the system development cycle, thereby minimizing the error cost~\cite{ErrorCE10, SDLC10}. Model checking can be summarized as an algorithmic technique for checking temporal properties of systems. Given a system $P$ and a temporal specification $\alpha$ on the runs of the system, decide whether system $P$ satisfies specification $\alpha$. This consists of checking that all runs of $P$ constitute models for $\alpha$. It suffices to show that no run of $P$ is a model for $\neg \alpha$, which is the same as checking that the intersection of the language accepted by $P$ and the language defined by $\neg \alpha$ is empty. For compex systems with millions of states, model checking can quickly run into the state space explosion problem~\cite{ClarkeKNZ11}. If there was a system with $n$ components and each component contained $m$ states, the composition of those components, would contain $m^n$ states, thereby exponentially incrementing the number of states. This issue is tackled, by performing model checking on \emph{bounded} (finite) runs of the system, and restricting the \emph{state space}. This technique is called \emph{bounded} model checking (BMC)~\cite{BCCZ99}. Here, the bound is chosen from heuristics or arbitrarily provided beforehand to the bounded model checker. Both decidable and undecidable logics have been used to perform BMC~\cite{LatvalaBHJ04,LatvalaBHJ05,TaoZCW07}. The third challenge is to automatically perform BMC for UCS.

In this paper, we make the following key contributions:
\begin{itemize}
    \item We model UCS using an extension of nets, called $\nu$-nets~\cite{VelardoF08} and describe a running example.
    \item We introduce a one-variable fragment of Monadic First Order Temporal Logic (MFOTL)~\cite{HWZ01} called {\Lstar} for expressing the properties.
    \item We provide the SMT encoding for {\Lstar} and $\nu$-nets in order to perform BMC. We also implement this in the tool UCSChecker~\cite{ucsChecker}.
\end{itemize}

\subsection*{Related Work}

In this section, we compare our work against other verification methods and existing literature.

\noindent\emph{Verification of unbounded nets:} In~\cite{AbdullaJ01}, they perform backwards reachability on unbounded Petri nets, however the tokens are not distinguishable. More recently~\cite{AmatDH22}, a semi-decision procedure to check the reachability of unbounded Petri nets is given, however this again does not consider distinguishable tokens. In~\cite{FelliGMRW23}, SMT has been used to perform conformance checking of data Petri nets, which is different from BMC. To the best of our knowledge, there is no prior work on verification of unbounded Petri nets with distinguishable tokens, and where the properties are expressed in some temporal logic.

\noindent\emph{Parametric verification:}
First we disambigute the notion of bounds. In the BMC approach, the number of steps upto which the system is verified, say $k$, indirectly places a bound on the number of tokens which are present in the net, thereby restricting the number of clients that are represented in the unbounded client-server system. For instance, when unrolling the system for $k$ steps in the BMC approach, it indirectly places a cap that there can be atmost $k$ newly generated tokens at a place (taking into consideration the weight of the arc and the initial marking).

In contrast to the BMC approach, there is existing literature on parametric verification such as~\cite{Lowe22}, where authors consider a fixed number of identifiable processes on which safety and deadlock-freedom is verified.
In~\cite{DecidableReasoningOOPSLA17}, they work with a decidable fragment of First Order Logic, with a
quantifier prefix $\exists^{\ast}\forall^{\ast}$, popularly known as the Bernays-Sch\"{o}nfinkel-Ramsey Class, where the temporal modalities are not considered in the same manner as in MFOTL. Their goal is to check the inductive invariants for safety and liveness properties, with expert inputs, whereas, ours focuses on automatic, push button verification of properties including safety, liveness and deadlock over unbounded $\nu$-nets without additional user intervention. Another key difference here, is how the fragment that we consider has only one sort, whereas, in~\cite{DecidableReasoningOOPSLA17}, there are many sorts, some of which are bounded due to domain knowledge.
In comparison with~\cite{KVW17}, where parametric verification of threshold automata where there are identifiers, our approach allows for an unbounded number of identified processes, without imposing an explicit bound on the number of processes.
In~\cite{LeaderElectionTACAS17,Lowe22,UnboundedMAS16,KVW17} the bounds are predetermined and in~\cite{DecidableReasoningOOPSLA17}, a part of the parameters are predetermined and bounded.

\noindent\emph{Interactive Theorem Proving:} While employing Interactive Theorem Provers or proof systems such as $TLA^+$~\cite{Chen2016,Chand2016}, user expertise and guidance during the verification process, in contrast to BMC approach which is automatic. Recently, in~\cite{KarpMillerCOQ17,KarpMillerMinCovCOQ24}, the interactive theorem prover COQ has been used for verifying the foundational Karp-Miller algorithm for computing the coverability tree. This technique has not been extended yet, to verify Petri net properties.

\subsection*{Outline}

In Sec.~\ref{sssec:model}, we describe the running example of UCS and identify the suitable formal model among the various extensions of nets. The syntax, semantics of logic {\Lstar} are in Sec.~\ref{sec:mlogic} and the bounded semantics necessary for performing BMC are in Sec.~\ref{ssec:bsem}. Based on the bounded semantics, we give the SMT encoding in Sec.~\ref{sec:propenc} and describe the BMC tool in Sec.~\ref{ssec:verif} and draw conclusions in Sec.~\ref{sec:concl}.
\section{Modeling unbounded client-server systems}\label{sssec:model}
We consider as a running example, the Autonomous Parking System (APS) that manages parking lots through communication between the system (server) and the vehicle (client). This system has been successfully implemented by the industry~\cite{spark11,CaiZQZD21}. This is a type of single server multiple client system, where the clients are distinguishable and unbounded. The service being offered is the finite set of parking lots available for occupancy by the clients. In this section, the objective is to identify a formal model for the combined interactions between the server and unboundedly many clients in the running example. The formal model should not be specific to the APS case study however, it needs to be generic enough to be applicable to other unbounded client-server systems as well.
In Sec.~\ref{ssec:casestudytravel}, we provide the travel agency case study to demonstrate this.

\begin{figure}[!ht]
    \begin{minipage}{0.5\textwidth}
        \centering
        \vspace{-10pt}
        \scalebox{0.6}{\begin{tikzpicture}[scale=0.4]

	\node 	(s1)		at	(0,10) 	[ellipse, thick, draw=red!50,fill=blue!20,inner sep=0pt,minimum height=.70cm, minimum width=1.5cm]{\small server\_ready(SR)};

	\node 	(s0)		at	(-12,10) 	[ellipse, thick, draw=red!50,fill=blue!20,inner sep=0pt,minimum height=.70cm, minimum width=1.5cm]{\small server\_busy(SB)};

	\node 	(s2)		at	(-6,6) 	[ellipse, thick, draw=red!50,fill=green!20,inner sep=0pt,minimum height=.70cm, minimum width=1.5cm]{\small request\_granted(RG)};

	\node 	(s3)		at	(-6,2) 	[ellipse, thick, draw=red!50,fill=green!20,inner sep=0pt,minimum height=.70cm, minimum width=1.5cm]{\small deallocate\_parking\_lot(DP)};

	\node 	(s4)		at	(6,6) 	[ellipse, thick, draw=red!50,fill=blue!20,inner sep=0pt,minimum height=.70cm, minimum width=1.5cm]{\small request\_rejected(RR)};

	\draw[->, bend left=10,thick]	(s0)	to	node[auto]{\small completed\_processing}	(s1);

	\draw[->, bend left=10, thick]	(s1)	to	node[auto]{\small processing}	(s0);

	\draw[->, thick]	(s1)	to	node[auto]{\small accept}	(s2);

	\draw[->, thick]	(s1)	to	node[auto,swap]{\small reject}	(s4);

	\draw[->, thick]	(s2)	to	node[auto]{\small successful vehicle exit}	(s3);

	\draw[->, thick, bend right = 10]	(s4)	to	node[auto,swap]{\small unsuccessful vehicle exit}	(s1);

\end{tikzpicture}}
        \caption{State diagram of server}
        \label{fig:system-model}
        \vspace{-10pt}
    \end{minipage}\hspace{1em}
    \begin{minipage}{0.5\textwidth}
        \centering
        \vspace{-10pt}
        \scalebox{0.6}{\begin{tikzpicture}[scale=0.4]

	\node 	(s1)		at	(0,10) 	[ellipse, thick, draw=red!50,fill=green!20,inner sep=0pt,minimum height=.70cm, minimum width=1.5cm]{\small parking\_requested(PR)};

	\node 	(s2)		at	(-4,6) 	[ellipse, thick, draw=red!50,fill=green!20,inner sep=0pt,minimum height=.70cm, minimum width=1.5cm]{\small occupy\_parking\_lot(OP)};

	\node 	(s3)		at	(-6,2) 	[ellipse, thick, draw=red!50,fill=green!20,inner sep=0pt,minimum height=.70cm, minimum width=1.5cm]{\small exited\_successfully(ES)};

	\node 	(s4)		at	(7.5,3) 	[ellipse, thick, draw=red!50,fill=green!20,inner sep=0pt,minimum height=.70cm, minimum width=1.5cm]{\small parking\_unavailable(PU)};

	\node 	(s5)		at	(7.5,-2) 	[ellipse, thick, draw=red!50,fill=green!20,inner sep=0pt,minimum height=.70cm, minimum width=1.5cm]{\small exited\_unsuccessfully(EU)};

	\draw[->, thick]	(s1)	to	node[auto]{\small server accepts}	(s2);

	\draw[->, thick]	(s1)	to	node[auto]{\small server rejects}	(s4);

	\draw[->, thick]	(s2)	to	node[auto]{\small successful exit}	(s3);

	\draw[->, thick]	(s4)	to	node[auto]{\small unsuccessful exit}	(s5);

\end{tikzpicture}}
        \caption{State diagram of client}
        \label{fig:vehicle-model}
        \vspace{-10pt}
    \end{minipage}
\end{figure}

We begin by describing the state diagrams for the server and client given in Fig.~\ref{fig:system-model} and Fig.~\ref{fig:vehicle-model} respectively. Initially, the system is in the state \emph{server\_ready (SR)}, when it is ready to service the client.
We assume a steady inflow of parking requests.
When a client inquires about parking space, the client is in the \emph{parking\_requested (PR)} state.
The server non-deterministically chooses to either grant or reject the parking request based on local information such as space availability, the priority of incoming requests, etc. We assume two disjoint workflows for each scenario.
First, if the server accepts the request, the server is in \emph{request\_granted (RG)} state and simultaneously, the client goes to \emph{occupy\_parking\_lot (OP)} state. Eventually, the client gives up its allocated parking space, is in \emph{exit\_successfully (ES)} and simultaneously the server is in \emph{deallocate\_parking\_lot (DP)} state. This marks the successful exit of the client from the system.
Second, if the server rejects the request, the client is in
\emph{parking\_unavailable (PU)} state and the server is in \emph{request\_rejected (RR)} state.
The only option is for the client to exit.
After granting the request, the server can go to \emph{server\_busy (SB)} state. Theoretically, this description allows for an unbounded number of client requests to be processed by the server, albeit there may be limitations on the availability of the finite parking space. We assume that the autonomous parking system can reasonably guide the vehicle manoeuvres within the parking lot. Revisiting our objective, it is not difficult to observe that the combined interactions between the server and clients described above can be interleaved and formally modeled as a single Petri net.

We consider the specific setting of single server multiple client systems, with distinguishable clients. These distinguishable clients are represented using distinguishable tokens of the Petri net, i.e, tokens appended with identifiers~\cite{VelardoF07}. There are an unbounded number of clients and a fresh client identifier is issued whenever a new client enters the system. When clients exit, the identifiers need to be purged. We outline the requirements for the formal model and arrive at a suitable representation.

Petri nets are suitable to model the \emph{concurrent} behaviour of the clients and are also suitable to capture an unbounded number of clients. The places correspond to the local states of the client and server. We have a disjoint set of server places and client places. The combined interactions of the server and client processes are represented by transitions. The tokens correspond to the processes (server process, client process). Unbounded Petri nets with indistinguishable tokens are not sufficient to differentiate between processes (server process, client process). Hence we look for another model.
At first glance, a candidate model is the colored Petri net (CPN)~\cite{JensenCPN07} which satisfies the above requirements. CPNs allow arbitrary expressions over user-defined syntax labelling the arcs, and the underlying modeling language (such as CPN Modeling Language in  CPN Tools~\cite{JensenKW07}) is highly expressive. However, we do not prefer the CPN, for the following engineering reasons.

First, there is a dearth of tools to automatically \emph{unfold} unbounded colored Petri nets. Recall that the big picture is the automatic verification of unbounded client-server systems. Alternatively, suppose we represent the formal model as a CPN such as using CPN Tools, there are no existing tools that can automatically \emph{unfold} an \emph{unbounded} CPN created using CPN Tools. Existing tools can only unfold \emph{bounded} CPNs~\cite{Dal-Zilio20,BilgramJPST22}. The second candidate model is a type of $\nu$-net~\cite{VelardoF08}, which is a CPN defined over a system of component nets, which use a labelling function $\lambda$, to handle synchronization between multiple component nets of a larger net system.

We restrict the $\nu$-nets to a single component, providing a simplified
definition while doing away with the labelling function used in $\nu$-nets. The client behaviour can be represented as a state machine. Similarly the server behaviour can also be represented as a state machine.
In this representation, we describe the behaviour of the single server multiple client system as a single component.

We begin with some definitions that are necessary for describing the restricted $\nu$-net. Given an arbitrary set $A$, we denote by $\mathcal{MS}(A)$, the set of finite multisets of A, given by the set of mappings $m:~A\to \mathbb{N}$. We denote by $S(m)$ the support of m, defined as follows:
$S(m)=\{a\in A \mid m(a)>0\}$. Distinguishable tokens (identifiers) are taken from an arbitrary infinite set \emph{Id}. To handle this, we add matching variables labeling the arcs, taken from a set $Var$. To handle the movement of tokens inside the $\nu$-net, we employ a finite set of variables, $Var$ using which we label the arcs of the net.

\begin{figure}[ht]
    \begin{center}
        \scalebox{0.8}{\usetikzlibrary {petri,positioning}
\begin{tikzpicture}[scale=0.7]
	\centering

	\node[transition,fill=black,minimum height=.4cm,minimum width=.4cm,label=left:{\small $t_{src}$}] (t0) {};

	\node[place,draw=red!80,label=right:{\small $p_{PR}$},minimum height=.70cm, minimum width=.5cm,below=of t0] (p0) {}	edge[pre] node[auto] {$\nu$} (t0);
	\node[token,fill = white, minimum height=0.40cm, minimum width=.5cm,text=black] at (p0) {\text{$\{1,2\}$}};

	\node[place,draw=blue!80,label=left:{\small $p_{SR}$},minimum height=.70cm, minimum width=.5cm,left=of p0] (p1){};
	\node[token,fill = white, text=black] at (p1) {$\{0\}$};

	\node[transition,fill=black,minimum height=.4cm,minimum width=.4cm,label=left:{\small $t_{acc}$}, below=of p1] (t1){}
	edge[pre] node[auto] {s} (p1)
	edge[post] node[auto] {} (p1)
	edge[pre] node[xshift=-5pt, yshift=-9pt] {c} (p0);

	\node[place,draw=red!80,label=left:{\small $p_{OP}$},minimum height=.70cm, minimum width=.5cm,below=of t1] (p2){}
	edge[pre] node[auto] {c} (t1) ;

	\node[transition,fill=black,minimum height=.4cm,minimum width=.4cm,label=left:{$t_{s\_exit}$},below=of p2] (t2){}
	edge[pre] node[auto] {c} (p2);

	\node[place,draw=red!80,label=left:{\small $p_{ES}$},minimum height=.70cm, minimum width=.5cm,below=of t2] (p4){}
	edge[pre] node[auto] {c} (t2);

	\node[transition,fill=black,minimum height=.4cm,minimum width=.4cm,label=left:{\small $t_{acc\_sink}$},below=of p4] (t3) {} edge[pre] node[auto] {c} (p4) ;

	\node[transition,fill=black,minimum height=.4cm,minimum width=.4cm,label=right:{\small $t_{rej}$},below=of p0] (t4){}
	edge[pre] node[xshift=-7pt,yshift=12pt] {s} (p1)
	edge[post] node[auto] {} (p1)
	edge[pre] node[auto] {c} (p0);

	\node[place,draw=red!80,label=right:{\small $p_{PU}$},minimum height=.70cm, minimum width=.5cm,below=of t4] (p6){}
	edge[pre] node[auto] {c} (t4) ;

	\node[transition,fill=black,minimum height=.4cm,minimum width=.4cm,label=right:{$t_{u\_exit}$},below= of p6] (t5) {}
	edge[pre] node[auto] {c} (p6);

	\node[place,draw=red!80,label=right:{\small $p_{EU}$},minimum height=.70cm, minimum width=.5cm,below=of t5] (p7){}
	edge[pre] node[auto] {c} (t5) ;

	\node[transition,fill=black,minimum height=.4cm,minimum width=.4cm,label=right:{$t_{rej\_sink}$},below= of p7] (t5) {}
	edge[pre] node[auto] {c} (p7);
\end{tikzpicture}}
        \caption{A $\nu$-net modeling APS}
        \label{fig:APS}
    \end{center}
\end{figure}
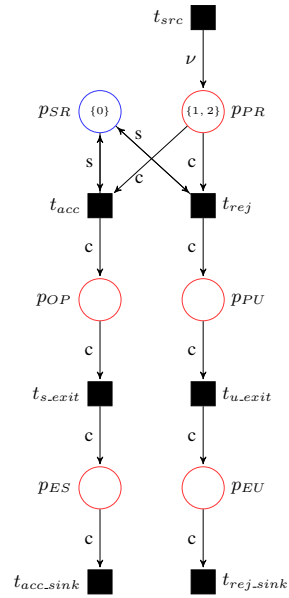

\begin{definition}\label{defn:nunet}
    A $\nu$-net is a coloured Petri net $N=(P,T,F)$, where
    \begin{itemize}
        \item $P$ and $T$ are finite disjoint sets of places and transitions, respectively,
        \item $F$: $(P \times T) \cup (T \times P) \to \mathcal{MS(\text{Var})}$ defines the
              set of arcs of the net, satisfying $\nu \not \in pre(t)$ for every $t
                  \in T$.
    \end{itemize}

    For a transition $t$ of the net, we define, $post(t)=\bigcup_{p\in P} S(F(t,p))$,
    $pre(t)=\bigcup_{p\in P} S(F(p,t))$
    and $Var(t)= pre(t) \bigcup post(t)$.

\end{definition}

For instance, in Fig.~\ref{fig:APS}, a set of transitions $T=$ $\{t_{acc}$, $t_{rej}$, $t_{s\_exit}$, $t_{u\_exit}$, $t_{acc\_sink}$, $t_{rej\_sink}\}$ and places $P=\{$ $p_{PR}$, $p_{SR}$, $p_{OP}$, $p_{PU}$, $p_{ES}$, $p_{EU}\}$. $pre(t_{acc})=\{s,c\}$ and $post(t_{acc})=\{s,c\}$, hence $Var(t_{acc})=\{s,c\}$.

\begin{definition}[Marking]
    A marking  of a restricted $\nu$-net $N=(P,T,F)$ is a function $M:P\to(\mathcal{MS}(Id))$.
\end{definition}
The $\nu$-net shown in Fig.~\ref{fig:APS}, has the initial marking $M_0=\langle \{0\}, \{1,2\}, \emptyset, \emptyset, \emptyset, \emptyset \rangle$.

\begin{definition}[Mode]
    We denote by $S(M)$ the set of identifiers in $M$. that is, $S(M)=\bigcup_{p\in P}S(M(p))$. A mode of a transition $t$ is a mapping $\sigma:Var(t)\to Id$, instantiating every variable in the adjacent arcs of $t$ to some identifier.
\end{definition}

Let $N$ be a restricted $\nu$-net and $M$ a marking of $N$ according to Defn.~\ref{defn:nunet}.

\begin{definition}[Enabling Rule]
    We say that $M$ \textbf{enables} the transition $t$ with mode $\sigma$ whenever:

    \begin{itemize}
        \item If $\nu \in Var(t)$ then $\sigma(\nu)\not \in S(M)$ and
        \item $\sigma(F(p,t))\subseteq M(p)$ for all $p \in P$.
    \end{itemize}
\end{definition}

Notice that if $\sigma(\nu)\not \in S(M)$ for the enabling of transition, that causes the creation of fresh (equal) identifiers in all the places reached by arcs labelled by the special variable $\nu \in Var$ that appears only in post-condition arcs.

\begin{definition}[Firing Rule]\label{nufire}
    The reached marking of net $N$ after firing of $t$ with mode $\sigma$
    is denoted by $M \xrightarrow{t(\sigma)}M'$, where
    $\forall p \in P: M'(p)=M(p) - \sigma(F(p,t)) + \sigma(F(t,p))$.
\end{definition}

For the $\nu$-net in Fig.~\ref{fig:APS} with initial marking $M=\langle \{0\}, \{1,2\}, \emptyset, \emptyset, \emptyset, \emptyset \rangle$ and $M \xrightarrow{t_{acc}} M'$, where $M'=\langle \{0\}, \{2\}, \{1\}, \emptyset, \emptyset, \emptyset \rangle$, i.e, the token $\{1\}$ has moved to the place $P_{OP}$. The mode is represented in the figure.

In Fig.~\ref{fig:APS}, the transitions $t_{acc}$, $t_{rej}$, $t_{s\_exit}$, $t_{u\_exit}$, $t_{acc\_sink}$, $t_{rej\_sink}$, represent the accept, reject, exit successfully, exit unsuccessfully and the two sink transitions respectively. A transition is \textbf{identifier-preserving} if $post(t)\setminus \{\nu\}\subseteq pre(t)$. Here, all of them are \emph{identifier-preserving} transitions, which ensures that the system with identified clients is represented correctly modeled. The firing of transition $t_{src}$ acts as the source. The arc labelled $\nu$ ensures that a new client identifier is generated in place $p_{PR}$. The place $p_{PR}$ contains a set of clients requesting for parking.
In the unsuccessful scenario, the transition $t_{rej}$ is fired when the server rejects the request, which brings the vehicle to \emph{parking\_unavailable} state represented by place $p_{PU}$. On firing of transition $t_{u\_exit}$, the vehicle goes to \emph{exited\_unsuccessfully} state represented by place $p_{EU}$.
The firing of transition $t_{rej\_sink}$ is the sink transition for the rejected parking requests. This ensures that the rejected vehicle identifier exits the system and is never reused. If the client arrives after it has exited, it is always issued a fresh identifier.
Notice that there are arcs labelled $s$ to indicate the server which
has identifier $0$, which is necessary for the acceptance or rejection
of a parking request. The token with identifier $0$ is permanently present in each marking
exactly at server place $p_{SR}$. The $\nu$-labelled arc ensures that new
identifiers are generated, essentially giving an unbounded number of agents in the $\nu$-net. The $c$-labelled ($s$-labelled) arcs carry the client
identifiers (server process) from one client place (server place) to another. The net behaves as a
standard $\nu$-net component with autonomous transitions as described in~\cite{VelardoF08}.
\section{The Monodic Logic {\Lstar}}\label{sec:mlogic}

The monodic logic {\Lstar} is an extension of Linear Temporal Logic (LTL), and both a syntactic and semantic subclass of MFOTL~\cite{HWZ01}. A \emph{monodic} formula is a well-formed formula with at most one free variable in the scope of a temporal modality. In this section, we describe the syntax and semantics of {\Lstar} with suitable examples. To give a flavour of {\Lstar} and its expressibility, we enumerate some properties of APS that are not easily expressible in Linear Temporal Logic (LTL).
Let $P_s$ be the set of atomic propositions of the server and $ P_c$ be the set of client predicates. In the APS running example, they are defined as follows:
\begin{align*}
  P_c
   & =\{parking\_requested (PR),~occupy\_parking\_lot (OP),     \\
   & \qquad~parking\_unavailable (PU), exit\_successfully (ES), \\
   & \qquad~exited\_unsuccessfully (EU)\}                       \\
  P_s
   & =\{server\_ready (SR)\}
\end{align*}

\begin{example}\label{ex:forall}
  When a vehicle requests a parking space, it is always the case that for every vehicle, it eventually exits the system, either successfully after being granted a parking space, or unsuccessfully, when its request is denied.
  \begin{align*}
    \psi_1 & =\mathbf{G}_s(\forall x) \Big( parking\_requested(x) \Rightarrow               \\
           & \qquad\mathbf{F}_c~ (exit\_successfully(x)~\lor~exit\_unsuccessfully(x)) \Big)
  \end{align*}
\end{example}

\begin{example}
  It is always the case that if the client occupies a parking lot, it will eventually exit the parking lot.
  \begin{align*}
    \psi_2 & =\mathbf{G}_s(\forall x) \Big(occupy\_parking\_lot(x)\Rightarrow \mathbf{F}_c(exit\_successfully(x))\Big)
  \end{align*}
\end{example}

\begin{example}
  There may be clients whose requests are rejected.
  \begin{align*}
    \psi_{3} & =\mathbf{G}_s(\exists x) \big(parking\_requested(x) \land \mathbf{F}_c (exit\_unsuccessfully(x))\big)
  \end{align*}
\end{example}

\begin{example}
  There may be clients who have requested for parking and who wait in the parking unavailable state until they are able to exit the system.
  \begin{align*}
    \psi_4 & = \mathbf{G}_s(\exists x) \bigg( parking\_requested(x)~\land                                      \\
           & \qquad \mathbf{F}_c \big( parking\_unavailable(x)~\mathbf{U}_c~exit\_unsuccessfully(x)\big)\bigg)
  \end{align*}
\end{example}

It can be observed that there are no free variables in the scope of $\mathbf{G_s}$ and exactly one free variable in the scope of the client modalities. It is also possible to construct {\Lstar} specifications with propositions from $P_s$ and server transitions.
The ease of expressibility of the client and server behaviour is the key motivation behind the logic {\Lstar} which is formally described in the subsequent section.

\begin{example}
  Recollect Ex.~\ref{ex:forall}, where we have
  \begin{align*}
    \psi_1 & =\mathbf{G}_s(\forall x) \Big( parking\_requested(x) \Rightarrow               \\
           & \qquad\mathbf{F}_c~ (exit\_successfully(x)~\lor~exit\_unsuccessfully(x)) \Big)
  \end{align*}

  Given that we know apriori, the set of clients in the system and we thereby have atomic client predicates, with subscript indicating the index of the client, we have the following equivalent LTL formula for $\psi_1$

  \begin{align*}
    \psi_1^{'} & =\mathbf{G} \Big( (parking\_requested_{0} \Rightarrow                          \\
               & \qquad\mathbf{F}~ (exit\_successfully_{0}~\lor~exit\_unsuccessfully_{0}))      \\
               & \land \qquad(parking\_requested_{1} \Rightarrow                                \\
               & \qquad\mathbf{F}~ (exit\_successfully_{1}~\lor~exit\_unsuccessfully_{1}))      \\
               & \land \qquad (parking\_requested_{2} \Rightarrow                               \\
               & \qquad\mathbf{F}~ (exit\_successfully_{2}~\lor~exit\_unsuccessfully_{2}))\Big)
  \end{align*}
\end{example}
\emph{Remark: }
First, the LTL formula $\psi_1^{'}$, has more clauses than the equivalent {\Lstar} formula $\psi_1$. While performing bounded model checking, metrics such as the number of clauses and the formula length plays a role in the time taken to verify. Second, we do not know the number of clients in UCS apriori, which is a bottle neck towards expressing formulas of UCS in LTL. As we have seen, we can express these properties in a straightforward manner using {\Lstar}.

In the upcoming section, we shall formally describe the syntax and semantics of {\Lstar}.

\subsection{Syntax of {\Lstar}}

The set of \emph{client formulae} $\Delta$, is the \emph{boolean and temporal modal closure} of atomic client formulae $P_c$:

\[\alpha,\beta \in \Delta ::= p(x), p\in P_c\mid \lnot\alpha \mid \alpha\lor\beta \mid  \alpha \land \beta \mid \mathbf{X}_c \alpha\mid \mathbf{F}_c\alpha \mid \mathbf{G}_c \alpha \mid \alpha ~\mathbf{U}_c~ \beta\]
The set of \emph{server formulae}, $\Psi$, is the \emph{boolean and temporal modal closure} of $\Phi=\{(\exists x)\alpha,(\forall x)\alpha \mid \alpha \in \Delta\}$ and atomic server formulae $P_s$:
\[\Psi ::= q \in P_s \mid \lnot \psi \mid \phi \in \Phi \mid \psi_1 \lor \psi_2 \mid \psi_1\land \psi_2 \mid \mathbf{X_s} \psi \mid \mathbf{F_s} \psi \mid \mathbf{G_s}\psi \mid \psi_1 ~\mathbf{U_s}~\psi_2\]

\noindent where $\psi,\psi_1,\psi_2 \in \Psi$.
It can be observed that in {\Lstar}, the quantifier depth is at most one and quantifier alternation is not allowed. The syntax allows us to specify only monodic formulas. Every variable in the server formulas is bounded; {\Lstar} allows for only client formulas to be quantified and we do not have quantifiers over server formulae.

\subsection{Semantics of {\Lstar}}\label{ssec:semantics}

\begin{figure}
  \centering
  \scalebox{0.7}{\input{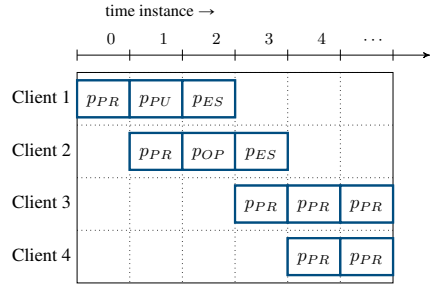}}
  \caption{Snapshot of the running example $(APS)$ depicting live windows}
  \label{fig:livewindows}

\end{figure}
We consider the \emph{unbounded client-server systems} where all clients are of the same type and they may enter and exit the system dynamically. At any point in time, the number of clients is \emph{bounded}, but their exact number is \emph{unknown} and \emph{dynamic}.
The clients which are currently in the system and have not yet exited are refered to as \emph{live clients} (\emph{live agents}).
This notion becomes particularly noticeable while introducing the identfiable clients (agents). The \emph{live window} of a particular client begins when it enters the system and ends when the client exits the system. Hence, if there are several \emph{live agents}, their \emph{live windows} would overlap each other. This is interesting as it allows us to reason about the \emph{live clients} which satisfy particular properties simultaneously. We illustrate these concepts with respect to the running example $APS$ before formally defining them.

\begin{example}
  Fig.~\ref{fig:livewindows} depicts the snapshot of the system with $4$ distinguishable clients, with overlapping \emph{live windows}, with the bound $5$. The $x$ axis denotes the time instance. The clients are along the $y$ axis.  While the system is \emph{unbounded}, there are a finite number of clients at an instant, as shown in this figure.
  Each row shows the local state of that client. For each instance, the local state of the client is in the cell i.e, client $1$ is at state $p_{PR}$ at instance $0$.
  For client $1$, the \emph{left boundary}, when it enters the system is at instance $0$ and its \emph{right boundary} is at instance $2$, when it exits the system . This corresponds to the client requesting parking and getting rejected.
  There may be multiple clients in the same local state (client $3$ and client $4$ are in state $p_{PR}$ at instance $4$). There may be clients which are live at the bound $5$ and have not exited the system, such as clients $3$ and $4$. This is an interesting case, where the bound (in the snapshot) is equal to the current right boundary for the client.
\end{example}

\noindent At the outset, we define the following objects:
\begin{itemize}
  \item Let $CN$ be a countable set of client names. The client enters the system and gets a unique identifier (name) from $CN$. When the client exits the system, the identifier gets discarded and is never reused.
  \item Let $CS = (Q,\Sigma,\delta,I,F)$ be a finite state machine describing the behaviour of a client. Let $L:Q \to 2^{P_c}$ be the definition (labelling function) of each client state  $q \in Q$ in terms of a subset of properties $P_c$ true in that state. The states in $F$ are sink states with no outgoing arcs.
  \item Let $\mathfrak{Z}:CN \times \mathbb{N}_0 \to Q$ be a partial mapping describing the local state of each client $a \in CN$ at an instance $i \in \mathbb{N}_0$. For instance, $\mathfrak{Z}(a,i)\in q$, means that the local state of each client $a$ at instance $i$ is state $q$, where $q \in Q$.
\end{itemize}

Formally, a model is a triple $M=(\nu,V,\xi)$ where
\begin{enumerate}
  \item $\nu$ gives the local behaviour of the server as follows:
        $\nu=\nu_0\nu_1\nu_2\ldots$, where for all $0 \le i$, $\nu_i \subseteq P_s$,
  \item $V$ gives the set of live agents (clients) at each instance.

        $V=V_0V_1V_2\ldots$, where for all $0 \le i$, $V_i$ is a finite subset of $CN$, gives the set of live agents at the $i$th instance.
        \newline
        For every $0\le i$, $V_i$ and $V_{i+1}$ satisfy  the following properties:
        \begin{enumerate}
          \item if $V_i \subseteq V_{i+1}$ then for every $a \in V_{i+1}-V_i$ such that $\mathfrak{Z}(a,i+1)\in I$.
          \item if $V_{i+1}\subseteq V_i$ then for every $a \in V_i-V_{i+1}$ such that $\mathfrak{Z}(a,i)\in F$.
        \end{enumerate}
        $V$ may satisfy the following interesting property. For any $a \in CN$ and $i \in \mathbb{N}_0$ if $a \in V_i$ and if there exists  $j >i$ such that $a \not \in V_j$ then we may define the left and right boundaries of the live window for $a$, denoted by $left(a)$ and $right(a)$ where $left(a)\le i\le right(a)$. If no such $j$ exists then there is no right boundary for the live window of that client.

  \item $\xi=\xi_0\xi_1\xi_2\cdots$, where for all $0 \le i$, $\xi_i:V_i\to 2^{P_c}$ gives the properties satisfied by each live agent at $i$th instance, in other words, the corresponding states of live agents. In terms of $\mathfrak{Z}$, $L(\mathfrak{Z}(a,i))=\xi_i(a)$. Alternatively, $\xi_i$ can be given as $\xi_i:V_i\times P_c \to\{\top,\bot\}$.
\end{enumerate}

\begin{example}In the running example (APS), the clients has exactly one initial place $I=parking\_requested~(PR)$ and two exiting places namely,
  $F=\{exit\_successfully~(ES)$, $ exit\_unsuccessfully~(EU)\}$.
\end{example}
We make use of the model described above, to provide the satisfiability relations for {\Lstar}.
\subsection{Satisfiability relations}\label{ssec:satrel}
Given a valid model $M=(\nu,V,\xi)$, the satisfiability relations for the logic, $\models$  and $\models_x$ can be defined, via induction over the structure of the server formulae $\psi\in \Psi$, and client formulae $\alpha\in \Delta$, respectively. We selectively describe some of the important relations that provide insight into the others. The detailed satisfiability relations are in \emph{Appendix~\ref{sec:unboundedsem}}.

\begin{enumerate}
  \item[1.] $M,i\models q$ iff $q \in \nu_i$.
  \item[] The atomic server formula $q \in P_s$  is satisfied in the model $M$ at instant $i$ if and only if $q$ is a member of the local state of the server $\nu_i$ at instant $i$, where $i \in \mathbb{N}_0$.

  \item[3.] $M,i \models (\exists x)\alpha$ iff $\exists a\in CN$, $a \in V_i$ and $M,[x\mapsto a],i\models_x \alpha$.
  \item[] The client formula $\alpha$ is satisfied at some instant $i$ if and only if $\alpha$ is satisfied in any of the \emph{live} clients in $V_i$ i.e, for some client $a \in CN$ and the client name $a \in V_i$, where $i \in \mathbb{N}_0$.

  \item[8.] $M,i\models \mathbf{F_s}\psi$ iff $\exists j \ge i$, $M,j\models \psi$.
  \item[] The server formula $\mathbf{F_s}\psi$ is satisfied at some instant $i$ if and only if there is some instant $j \ge i$, where the formula $\psi$ holds.

  \item[16.] $M,[x\mapsto a],i\models_x \mathbf{F}_c\alpha$ iff $\exists j\ge i$, $a\in V_j$, $M,[x\mapsto a],j\models_x \alpha$ .
  \item[] The client formula $\mathbf{F}_c\alpha$ is satisfied at some instant $i$ if and only if there is some instant $j \ge i$, where the formula $\alpha$ holds in any of the \emph{live} clients in $V_j$.
\end{enumerate}

\subsection{Observations}\label{ssec:obsr}
We list down some observations about the logic {\Lstar} which supplement our
understanding of the language.
\begin{definition}{Ghost Interval: }
  Given a temporal property of a client, whose right boundary is
  $\lambda '$ such that the $\lambda ' < \lambda$. There is a ghost interval between
  $\lambda$ and $\lambda'$ where the formula $\alpha$ need not be asserted.
\end{definition}

\begin{figure}[!ht]
  \begin{minipage}{0.5\textwidth}
    \centering
    \scalebox{0.6}{		\begin{tikzpicture}[shorten >=1pt,node distance=2cm,on grid,auto] 
		\node[state] (s)   {}; 
		\node[state,font=\Large] (si) [ right=of s] {$s_i$};
		\node[state] (si1) [ right=of si] {};
		\node[state,font=\Large] (sk) [ right=of si1] {$s_\lambda$};
		\path[->] 
		(s) edge  node {} (si)
		(si) edge  node {} (si1)
		(si1)	edge  node{} (sk);
		\end{tikzpicture}	}
    \caption{Bounded loop-free path of length $\lambda$}
    \label{fig:loopfreepath}
  \end{minipage}\hspace{1em}
  \begin{minipage}{0.5\textwidth}
    \centering
    \scalebox{0.6}{\begin{tikzpicture}[shorten >=1pt,node distance=2cm,on grid,auto,minimum size=1em]
	\node[state] (s)   {};
	\node[state,font=\Large] (sl) [ right=of s] {$s_l$};
	\node[state] (sl1) [ right=of sl] {};
	\node[state,font=\Large] (si) [ right=of sl1] {$s_i$};
	\node[state,font=\Large] (sk) [ right=of si] {$s_\lambda$};
	\path[->]
	(s) edge  node {} (sl)
	(sl)	edge  node{} (sl1)
	(sl1)	edge  node{} (si)
	(si)	edge  node{} (sk)
	(sk)	edge [bend right] node {}(sl)	;
\end{tikzpicture}}
    \caption{Bounded path with $(\lambda,l)$ - loop}
    \label{fig:klloop}
  \end{minipage}
  \vspace{-1em}
\end{figure}

In Fig.~\ref{fig:livewindows}, there is a \emph{ghost interval} for client $1$ between $1$ (right boundary) and $5$ (the bound $\lambda$).\\

Let $\psi_5 = (\forall x) \Big( \mathbf{G}_c \big(parking\_requested(x)
    \mathbf{U}_c~ exit\_successfully(x) \big)\Big)$. In $\psi_5$,
$\mathbf{G}_c$ holds only till the \emph{live window} ends for the corresponding client i.e, this will be asserted only as long as the client is \emph{live}. Similarly, all other client temporal modalities $\mathbf{F}_c$, $\mathbf{U}_c$, and $\mathbf{X}_c$ need not be asserted beyond the \emph{live window} of that particular client.

At any given time, there are as many client lassos as there are clients, which are then combined to
construct the server lasso. For currently \emph{active} clients the right boundary
will be the execution bound $\lambda$.

We list down some of the properties that cannot be expressed in {\Lstar}:
\begin{enumerate}
  \item Consider the property, where there is at least one token in either place $p_{SR}$ or $p_{SB}$ or $p_{RR}$. From the net, $p_{SR}+p_{SB}+p_{RR} = 1$ is an invariant.
        These places represent the server being available, busy and the server rejecting the request.

  \item The following property is not expressible in {\Lstar}: It is always the case that the client request with an identifier (say) $1001$ is rejected.
  \item It is always the case that if the client request is accepted at time instance $x$, then the client exits the parking space at time instance $z$, $z>x$.

  \item Formulas with quantifiers before server temporal modality are not permitted.

  \item We do not have formulas of type, client $i$ satisfies a formula whereas client $j$ doesn't satisfy the same formula. We do not have equality/inequality to differentiate between the clients. We assume that all clients behave the same way and are of the same type. In the future, this model could be extended to clients of a finite set of types.

  \item We cannot compare clients in the same state.

        \begin{align*}
          \psi_6 = \mathbf{G}_s ((\exists x)(\exists y)parking\_requested(x) \land parking\_requested(y) \land \\
          \mathbf{F}_c(exit\_successfully (x) \land \mathbf{F}_c (exit\_successfully (y)))
        \end{align*}

  \item Here, there are no free variables either in the scope of $\mathbf{G_s}$ or $\mathbf{F_s}$.
        \begin{align*}
          \text{Let } \psi_7 = \big(\mathbf{G_s} (\exists x)~parking\_requested(x)\big) \land \big(\mathbf{F_s}(\exists y)~parking\_requested(y)\big)
        \end{align*}
  \item[]  The formula $\psi_7$ states that it is always the case that there is a client who has requested parking and eventually, there exists a client who has also requested parking. However, we cannot compare these two clients in our logic.
\end{enumerate}
In the next section, we arrive at the bounded semantics of {\Lstar} which are necessary for performing BMC.
\section{Bounded Semantics}\label{ssec:bsem}

In this section, we describe the bounded semantics of {\Lstar} in order to arrive at the SMT encoding, which is necessary for BMC.
We use $\models^{k}$ as a restriction on $\models$, where $k$ is the bound in the BMC strategy.
To tackle the unbounded number of clients as well as unboundedness in the unfolding of the model, it we introduce two bounds. The bounds are $\kappa$ and $\lambda$, where $\kappa$ is a bound on the number of clients and $\lambda$ is a bound on the execution steps of the net. The two parameters $\kappa$ and $\lambda$ are independent of each other. Since the runs are bounded, there are atmost $\kappa$ clients in the system, namely CN = $\{0,\ldots ,\kappa-1\}$.

First, we describe the bounded semantics without loop for a subset of formulas, as shown in Fig.~\ref{fig:loopfreepath} where $0\leq i \leq \lambda$, where $i$ is the current instance on the bounded path. Detailed semantics are in \emph{Appendix~\ref{sec:boundedsemboth}}.

\begin{figure}[!t]
    \begin{minipage}{0.4\textwidth}
        \centering
        \scalebox{0.7}{\input{live_windows_overlap_line}}
        \caption{Snapshot at bound = $2$}
        \label{fig:livewindowsline}
    \end{minipage}
    \begin{minipage}{0.6\textwidth}
        \centering
        \scalebox{0.7}{\begin{ganttchart}[
    vgrid,hgrid,
    progress label text=\relax,
    x unit=10mm,
    title height=1,
    foobar top shift=.15,
    foobar height=.70
  ]{0}{5}
  \textganttbar{Client 1}{$pc_1$}{0}{0}  \textganttbar{}{$pc_2$}{1}{1} \\
  \textganttbar{Client 2}{$pc_1$}{1}{1}  \textganttbar{}{$pc_3$}{2}{2}\\
  \textganttbar{Client 3}{$pc_1$}{3}{3} \textganttbar{}{$pc_1$}{4}{4}\textganttbar{}{$pc_1$}{5}{5}\\
  \textganttbar{Client 4}{$pc_1$}{4}{4} \textganttbar{}{$pc_4$}{5}{5}
  \begin{scope}[yshift=0.33cm]
    \draw[-] (0,0) -- (6,0);
    \draw ([yshift=-3pt]0,0) -- ++([yshift=6pt]0,0) node[xshift=45pt, yshift=15pt, above] {time instance $\rightarrow$};
    \draw ([yshift=-3pt]1,0) -- ++([yshift=6pt]0,0) node[xshift=-10pt, above] {$0$};
    \draw ([yshift=-3pt]2,0) -- ++([yshift=6pt]0,0) node[xshift=-10pt, above] {$1$};
    \draw ([yshift=-3pt]3,0) -- ++([yshift=6pt]0,0) node[xshift=-10pt, above] {$2$};
    \draw ([yshift=-3pt]4,0) -- ++([yshift=6pt]0,0) node[xshift=-10pt, above] {$3$};
    \draw ([yshift=-3pt]5,0) -- ++([yshift=6pt]0,0) node[xshift=-10pt, above] {$4$};
    \draw ([yshift=-3pt]6,0) -- ++([yshift=6pt]0,0) node[xshift=-10pt, above] {$5$};
  \end{scope}
\end{ganttchart}}
        \caption{Snapshot at bound = $5$ }
        \label{fig:livewindowsline3}
    \end{minipage}
\end{figure}

\begin{enumerate}
    \item[3.] $M,i \models^{k} (\exists x)\alpha$ iff $\exists a\in CN$, $a \in V_i$ and $M,[x\mapsto a],i\models^{k}_x \alpha$.

    \item[] The above formula is satisfied when there is at least one live client such that $\alpha$ is satisfied in the model at instance $i$ for the particular live client $a$.
        Given $\alpha={pc}_1 \lor {pc}_2$ and the snapshot of the system with the bound $2$ is in Fig.~\ref{fig:livewindowsline}. Here, $CN=\{1,2\}$. At instance $i=0$, model $M$ satisfies formula $\alpha$. Hence, this formula holds true at instance $0$.

    \item[4.] $M,i \models^{k} (\forall x)\alpha$ iff $\forall a\in CN$, if $a \in V_i$ then $M,[x\mapsto a],i\models^{k}_x \alpha$.

    \item[] The above formula is satisfied when the $\alpha$ is satisfied in the model at instance $i$ for all live clients in the set $CN$. Given $\alpha={pc}_4$ and Fig.~\ref{fig:livewindowsline}. The model $M$ does not satisfy the formula for all clients, since client $2$ does not satisfy $\alpha$.

    \item[8.] $M,i\models^{k} \mathbf{F_s}\psi$ iff $\exists j: i \le j \le\lambda$ , $M,j\models^{k} \psi$.

    \item[] This formula is satisfiable if there is some instance $j$ such that $i \le j \le\lambda$ at which the property $\psi$ holds. Given $\psi= (\exists x){pc}_4$. In Fig.~\ref{fig:livewindowsline}, the formula is unsatisfiable. However, in Fig.~\ref{fig:livewindowsline3}, it is satisfiable at instance $5$, where client $4$ is in local state ${pc}_4$.

    \item[16.] $M,[x\mapsto a],i\models^{k}_x \mathbf{F}_c\alpha$ iff $\exists j: i \le j \le\lambda$, $a\in V_j$ and $M,[x\mapsto a],j\models^{k}_x \alpha$.

    \item[] This formula is satisfiable if there is some instance $j$ such that $i \le j \le\lambda$ at which the property $\alpha$ is satisfied for the client $a$ and the client $a$ is a live client at instance $j$. Given $\alpha={pc}_2$. In Fig.~\ref{fig:livewindowsline} and Fig.~\ref{fig:livewindowsline3}, the formula is satisfiable at instance $1$ due to client $1$.

\end{enumerate}

Notice that in the bounded semantics equations $15$-$18$, the semantics depend crucially on whether $a$ is a live agent. Similarly, we can describe the bounded semantics with $(\lambda,l)$ - loop similar to the $(k,l)$ - loop in ~\cite{BiereCCSZ03}. As shown in Fig.~\ref{fig:klloop}, $i$ is the current instance on the bounded path, where the formula is evaluated, $\lambda$ is the bound and $l$ is the start position of the loop, where $0\leq i \leq \lambda$ and $0\leq l \leq \lambda$. The bounded semantics for this case are detailed in \emph{Appendix~\ref{ssec:boundedLoop}}. Consider the following lemma that relates the bounded and unbounded semantics of logic {\Lstar}.

The Lemma 1, states that if a LTL formula is satisfied in a bounded run, then it is also satisfied in an unbounded run.
\begin{lemma}[~\cite{BiereCCSZ03}]\label{lem:boundltlbiere}
    Let f be a LTL formula and $\Pi$ be a path then $\Pi \models^{k} f \implies \Pi\models f$.
\end{lemma}

Following this, we state a similar lemma for {\Lstar}:
\begin{lemma}\label{lem:bound}
    Let f be a {\Lstar} formula and $\Pi$ be a path then $\Pi \models^{k}_x f \implies \Pi\models_{x} f$.
\end{lemma}

MFOTL is decidable~\cite{HWZ01}. {\Lstar} is a syntactic fragment of MFOTL which is monodic and restricted to quantifier rank 1. Therefore, we have the following theorem:
\begin{theorem}
    {\Lstar} is decidable.
\end{theorem}

Now that the bounded semantics are established, we can give the SMT encoding of the logic and the model in the upcoming section.
\section{SMT Encoding for {\Lstar} and $\nu$-nets}\label{sec:propenc}

In this section, we first describe the SMT encoding for {\Lstar} which will enable us to implement a bounded model checker tool for restricted $\nu$-nets using {\Lstar} specifications. Let $[\mathcal{M}]_{\langle\lambda,\kappa\rangle}$ be the SMT encoding of $\lambda$-bounded runs of the net $\mathcal{M}$ containing at most $\kappa$ agents. The SMT encoding $[\mathcal{M}]_{\langle\lambda,\kappa\rangle}$ can be given based on the definition of the restricted $\nu$-net. Detailed encoding is in Appendix.~\ref{sec:smtnu}. Let $\psi$ be a property written in the logic language {\Lstar} which is model-checked in bounded runs of $\mathcal{M}$. For a given ${\langle\lambda,\kappa\rangle}$ and $i$ ($\psi$ is asserted at $i$, $0\le i\le \lambda$), we define two encoding functions $\prescript{}{}[\psi]^i_{\langle\lambda,\kappa\rangle}$ and $\prescript{}{l}[\psi]^i_{\langle\lambda,\kappa\rangle}$.
The formula $[\psi]^i_{\langle\lambda,\kappa\rangle}$ denotes the SMT encoding of $\psi$, where the bounded run of length $\lambda$ and at most $\kappa$ agents in the system do not contain a loop. The formula $\prescript{}{l}[\psi]^i_{\langle\lambda,\kappa\rangle}$ denotes the SMT encoding of $\psi$, where the bounded run of length $\lambda$ and at most $\kappa$ agents contains a loop which is asserted at $i$.

We add the following formulas about the dead agents (agents that are not alive) to the system specification:

\begin{itemize}
	\item If a client exits the parking lot, then it becomes dead in the next state.

	      $\delta_1 = \underset{{0\le i \le \lambda -1}}{\bigwedge} \bigg( \underset{{0\le \mathfrak{j} \le \kappa -1}}{\bigwedge}exit\_successfully(\mathfrak{j}, i) \To (dead(\mathfrak{j},i+1))\bigg)$

	      $\delta_2 = \underset{{0\le i \le \lambda -1}}{\bigwedge} \bigg( \underset{{0\le \mathfrak{j} \le \kappa -1}}{\bigwedge}exit\_unsuccessfully(\mathfrak{j},i) \To (dead(\mathfrak{j},i+1))\bigg)$

	\item If a client is dead then it remains dead.

	      $\delta_3 = \underset{{0\le i \le \lambda -1}}{\bigwedge} \bigg( \underset{{0\le \mathfrak{j} \le \kappa -1}}{\bigwedge}dead(\mathfrak{j}) \To (dead(\mathfrak{j},i+1))\bigg)$
\end{itemize}

\noindent In case of a different case study, where we consider any other net, instead of APS, we can suitably replace $exit\_successfully(\mathfrak{j})$, $exit\_unsuccessfully(\mathfrak{j})$ by the relevant termination condition(s).

The SMT encoding is derived from the bounded semantics in Sec.~\ref{sec:mlogic}. The following encodings are extensions of similar mappings defined in~\cite{BCCZ99}. First, we inductively define $[\psi]^i_{\langle\lambda,\kappa\rangle}$ as follows:
\begin{enumerate}
	\item $\prescript{}{}[q]^i_{\langle\lambda,\kappa\rangle} \equiv q_i$

	\item $\prescript{}{}[\lnot q]^i_{\langle\lambda,\kappa\rangle} \equiv \lnot q_i$

	\item $\prescript{}{}[(\exists x)\alpha]^i_{\langle\lambda,\kappa\rangle} \equiv  \underset{0\le \mathfrak{j} \le \kappa-1}{\bigvee}\prescript{}{}[\alpha[\mathfrak{j}/x]]^i_{\langle\lambda,\kappa\rangle}$

	\item $\prescript{}{}[(\forall x)\alpha]^i_{\langle\lambda,\kappa\rangle} \equiv \underset{0\le \mathfrak{j} \le \kappa-1}{\bigwedge}\prescript{}{}[\alpha[\mathfrak{j}/x]]^i_{\langle\lambda,\kappa\rangle}$

	\item $\prescript{}{}[\psi_1 \lor \psi_2]^i_{\langle\lambda,\kappa\rangle} \equiv \prescript{}{}[\psi_1]^i_{\langle\lambda,\kappa\rangle} \lor \prescript{}{}[\psi_2]^i_{\langle\lambda,\kappa\rangle}$

	\item $\prescript{}{}[\psi_1 \land \psi_2]^i_{\langle\lambda,\kappa\rangle} \equiv \prescript{}{}[\psi_1]^i_{\langle\lambda,\kappa\rangle} \land \prescript{}{}[\psi_2]^i_{\langle\lambda,\kappa\rangle}$

	\item $\prescript{}{}[\mathbf{X_s} \psi]^i_{\langle\lambda,\kappa\rangle} \equiv
		      \begin{cases} \prescript{}{}[\psi]^{i+1}_{\langle\lambda,\kappa\rangle} & \mbox{if ($i<\lambda$)}\\ False & \mbox{otherwise}\end{cases}$

	\item $\prescript{}{}[\mathbf{F_s}\psi]^i_{\langle\lambda,\kappa\rangle} \equiv \underset{i\le j \le \lambda}{\bigvee}\prescript{}{}[\psi]^{j}_{\langle\lambda,\kappa\rangle}$

	\item $\prescript{}{}[\mathbf{G_s} \psi]^i_{\langle\lambda,\kappa\rangle} \equiv False$

	\item $\prescript{}{}[\psi_1\mathbf{U_s}\psi_2]^i_{\langle\lambda,\kappa\rangle} \equiv \underset{i\le j \le \lambda}{\bigvee}([\psi_2]^{j}_{\langle\lambda,\kappa\rangle} \land \underset{i\le j' < j}{\bigwedge}[\psi_1]^{j'}_{\langle\lambda,\kappa\rangle})$

	\item 	$\prescript{}{}[p(\mathfrak{j})]^i_{\langle\lambda,\kappa\rangle} \equiv p(\mathfrak{j},i)$

	\item $\prescript{}{}[\lnot \alpha]^i_{\langle\lambda,\kappa\rangle} \equiv \lnot \prescript{}{}[\alpha]^i_{\langle\lambda,\kappa\rangle}$

	\item $\prescript{}{}[\alpha \lor \beta]^i_{\langle\lambda,\kappa\rangle} \equiv \prescript{}{}[\alpha]^i_{\langle\lambda,\kappa\rangle} \lor \prescript{}{}[\beta]^i_{\langle\lambda,\kappa\rangle}$

	\item $\prescript{}{}[\alpha \land \beta]^i_{\langle\lambda,\kappa\rangle} \equiv \prescript{}{}[\alpha]^i_{\langle\lambda,\kappa\rangle} \land \prescript{}{}[\beta]^i_{\langle\lambda,\kappa\rangle}$

	\item $\prescript{}{}[\mathbf{X}_c \alpha]^i_{\langle\lambda,\kappa\rangle} \equiv
		      \begin{cases} \big(dead(\mathfrak{j},i+1) \To False\big) \land \big(\lnot dead(\mathfrak{j},i+1) \To \prescript{}{}[\alpha]^{i+1}_{\langle\lambda,\kappa\rangle}\big) & \mbox{if ($i<\lambda$)} \\
              False                                                                                                                                                   & \mbox{if ($i=\lambda$)}\end{cases}$
	\item[] If the client is dead at the instance $i+1$, the original formula evaluates to false. If the client is live, the standard bounded semantics apply.
		Similar semantics are given in equations $16-18$ based on the client liveness at instance $j$.

	\item $\prescript{}{}[\mathbf{F}_c\alpha]^i_{\langle\lambda,\kappa\rangle} \equiv \underset{i \le j \le \lambda}{\bigvee}\bigg(\big(dead(\mathfrak{j},j) \To False\big) \land \big(\lnot dead(\mathfrak{j},j) \To \prescript{}{}[\alpha]^j_{\langle\lambda,\kappa\rangle}\big)\bigg)$

	\item $\prescript{}{}[\mathbf{G}_c\alpha]^i_{\langle\lambda,\kappa\rangle} \equiv \underset{i \le j \le \lambda}{\bigwedge}\bigg(\big(dead(\mathfrak{j},j) \To False\big) \land \big(\lnot dead(\mathfrak{j},j) \To \prescript{}{}[\alpha]^j_{\langle\lambda,\kappa\rangle}\big)\bigg)$

	\item $\prescript{}{}[\alpha\mathbf{U}_c\beta]^i_{\langle\lambda,\kappa\rangle} \equiv
		      \bigvee\limits_{i\leq j\leq \lambda}\bigg(\big(dead(\mathfrak{j},j) \To False\big) \land \big(\lnot dead(\mathfrak{j},j) \To {prop\_encode}_{\alpha\mathbf{U}_c\beta} \big)\bigg)$

	      \begin{itemize}
		      \item[] where ${prop\_encode}_{\alpha\mathbf{U}_c\beta}= \bigwedge\limits_{i\leq j\leq \lambda}\prescript{}{}[\beta]^j_{\langle\lambda,\kappa\rangle}
				      \land
				      \bigvee\limits_{i\leq j'< j}\prescript{}{}[\alpha]^{j'}_{\langle\lambda,\kappa\rangle}$
	      \end{itemize}

\end{enumerate}

Second, we inductively define $\prescript{}{l}[\psi]^i_{\langle\lambda,\kappa\rangle}$, for the loop case as follows:
\begin{enumerate}

	\item $\prescript{}{l}[q]^i_{\langle\lambda,\kappa\rangle} \equiv q_i$
	\item $\prescript{}{l}[\lnot q]^i_{\langle\lambda,\kappa\rangle} \equiv \lnot q_i$

	\item $\prescript{}{l}[(\exists x)\alpha]^i_{\langle\lambda,\kappa\rangle} \equiv  \underset{0\le \mathfrak{j} \le \kappa-1}{\bigvee}\prescript{}{l}[\alpha[\mathfrak{j}/x]]^i_{\langle\lambda,\kappa\rangle}$

	\item $\prescript{}{l}[(\forall x)\alpha]^i_{\langle\lambda,\kappa\rangle} \equiv \underset{0\le \mathfrak{j} \le \kappa-1}{\bigwedge}\prescript{}{l}[\alpha[\mathfrak{j}/x]]^i_{\langle\lambda,\kappa\rangle}$

	\item $\prescript{}{l}[\psi_1 \lor \psi_2]^i_{\langle\lambda,\kappa\rangle} \equiv \prescript{}{l}[\psi_1]^i_{\langle\lambda,\kappa\rangle} \lor \prescript{}{l}[\psi_2]^i_{\langle\lambda,\kappa\rangle}$

	\item $\prescript{}{l}[\psi_1 \land \psi_2]^i_{\langle\lambda,\kappa\rangle}\equiv \prescript{}{l}[\psi_1]^i_{\langle\lambda,\kappa\rangle} \land \prescript{}{l}[\psi_2]^i_{\langle\lambda,\kappa\rangle}$

	\item $\prescript{}{l}[\mathbf{X_s} \psi]^i_{\langle\lambda,\kappa\rangle} \equiv
		      \begin{cases} \prescript{}{l}[\psi]^{i+1}_{\langle\lambda,\kappa\rangle} & \mbox{if ($i<\lambda$)}\\ \prescript{}{l}[\psi]^l_{\langle\lambda,\kappa\rangle} & \mbox{if ($i=\lambda$)}\end{cases}$

	\item $\prescript{}{l}[\mathbf{F_s}\psi]^i_{\langle\lambda,\kappa\rangle} \equiv \underset{min(l,i)\le j \le \lambda}{\bigvee}\prescript{}{l}[\psi]^j_{\langle\lambda,\kappa\rangle}$

	\item $\prescript{}{l}[\mathbf{G_s} \psi]^i_{\langle\lambda,\kappa\rangle} \equiv \underset{min(l,i)\le j \le \lambda}{\bigwedge}\prescript{}{l}[\psi]^j_{\langle\lambda,\kappa\rangle}$

	\item $\prescript{}{l}[\psi_1\mathbf{U_s}\psi_2]^i_{\langle\lambda,\kappa\rangle} \equiv
		      \begin{cases}
			      \underset{i\le j \le \lambda}{\bigvee}(\prescript{}{l}[\psi_2]^{j}_{\langle\lambda,\kappa\rangle} \land \underset{i\le j' < j}{\bigwedge}\prescript{}{l}[\psi_1]^{j'}_{\langle\lambda,\kappa\rangle})                  & \mbox{if $(i\le l)$} \\
			                                                                                                                                                                                                                             &                      \\
			      \bigg( \big(\underset{i\le j \le \lambda}{\bigvee}(\prescript{}{l}[\psi_2]^{j}_{\langle\lambda,\kappa\rangle} \land \underset{i\le j' < j}{\bigwedge}\prescript{}{l}[\psi_1]^{j'}_{\langle\lambda,\kappa\rangle})\big) &                      \\
			      \hspace{25mm}\text{or}                                                                                                                                                                                                 & \mbox{if $(i> l)$}   \\
			      \big(\underset{l\le j < i}{\bigvee}(\prescript{}{l}[\psi_2]^{j}_{\langle\lambda,\kappa\rangle} \land \underset{l\le j' < j}{\bigwedge}\prescript{}{l}[\psi_1]^{j'}_{\langle\lambda,\kappa\rangle}) \big) \bigg)        &                      \\
		      \end{cases}$

	\item $\prescript{}{l}[p(\mathfrak{j})]^i_{\langle\lambda,\kappa\rangle} \equiv p(\mathfrak{j},i)$

	\item $\prescript{}{l}[\lnot \alpha]^i_{\langle\lambda,\kappa\rangle} \equiv \lnot \prescript{}{l}[\alpha]^i_{\langle\lambda,\kappa\rangle}$

	\item $\prescript{}{l}[\alpha \lor \beta]^i_{\langle\lambda,\kappa\rangle} \equiv \prescript{}{l}[\alpha]^i_{\langle\lambda,\kappa\rangle} \lor \prescript{}{l}[\beta]^i_{\langle\lambda,\kappa\rangle}$

	\item $\prescript{}{l}[\alpha \land \beta]^i_{\langle\lambda,\kappa\rangle} \equiv \prescript{}{l}[\alpha]^i_{\langle\lambda,\kappa\rangle} \land \prescript{}{l}[\beta]^i_{\langle\lambda,\kappa\rangle}$

	\item $\prescript{}{l}[\mathbf{X}_c \alpha]^i_{\langle\lambda,\kappa\rangle} \equiv
		      \begin{cases}
			      \big(dead(\mathfrak{j},i+1) \To False\big) \land \big(\lnot dead(\mathfrak{j},i+1) \To \prescript{}{l}[\alpha]^{i+1}_{\langle\lambda,\kappa\rangle}\big) & \mbox{if ($i<\lambda$)} \\

			      \big(dead(\mathfrak{j},l) \To False\big) \land \big(\lnot dead(\mathfrak{j},l) \To \prescript{}{l}[\alpha]^l_{\langle\lambda,\kappa\rangle}\big)         & \mbox{if ($i=\lambda$)}
		      \end{cases}$

	\item[] If the client is dead at the next instance ($i+1$ or $l$, respectively), the original formula evaluates to false. If the client is live, the standard bounded semantics apply. Similar semantics are given in equations $16-18$ based on the client liveness at instance $j$.

	\item $\prescript{}{l}[\mathbf{F}_c\alpha]^i_{\langle\lambda,\kappa\rangle} \equiv \underset{min(l,i) \le j \le \lambda}{\bigvee}\bigg(\big(dead(\mathfrak{j},j) \To False\big)  \land \big(\lnot dead(\mathfrak{j},j) \To \prescript{}{l}[\alpha]^j_{\langle\lambda,\kappa\rangle}\big)\bigg)$

	\item $\prescript{}{l}[\mathbf{G}_c \alpha]^i_{\langle\lambda,\kappa\rangle} \equiv \underset{min(l,i) \le j \le \lambda}{\bigwedge}\bigg((dead(\mathfrak{j},j) \To False) \land (\lnot dead(\mathfrak{j},j) \To \prescript{}{l}[ \alpha]^j_{\langle\lambda,\kappa\rangle})\bigg)$

	\item $\prescript{}{l}[\alpha\mathbf{U_c}\beta]^i_{\langle\lambda,\kappa\rangle} \equiv \\
		      \underset{min(l,i) \le j \le \lambda}{\bigvee}\bigg(\big(dead(\mathfrak{j},j) \To False\big) \land \big(\lnot dead(\mathfrak{j},j) \To {loop\_prop\_encode}_{\alpha\mathbf{U_c}\beta} \big)\bigg)$
	      ${loop\_prop\_encode}_{\alpha\mathbf{U_c}\beta}=
		      \begin{cases}
			      \underset{i\le j \le \lambda}{\bigvee}(\prescript{}{l}[\beta]^{j}_{\langle\lambda,\kappa\rangle} \land \underset{i\le j' < j}{\bigwedge}\prescript{}{l}[\alpha]^{j'}_{\langle\lambda,\kappa\rangle})                  & \mbox{if $(i\le\lambda)$} \\
			                                                                                                                                                                                                                            &                           \\
			      \bigg( \big(\underset{i\le j \le \lambda}{\bigvee}(\prescript{}{l}[\beta]^{j}_{\langle\lambda,\kappa\rangle} \land \underset{i\le j' < j}{\bigwedge}\prescript{}{l}[\alpha]^{j'}_{\langle\lambda,\kappa\rangle})\big) &                           \\
			      \hspace{25mm}\text{or}                                                                                                                                                                                                & \mbox{if $(i>\lambda)$}   \\
			      \big(\underset{l\le j < i}{\bigvee}(\prescript{}{l}[\beta]^{j}_{\langle\lambda,\kappa\rangle} \land \underset{l\le j' < j}{\bigwedge}\prescript{}{l}[\alpha]^{j'}_{\langle\lambda,\kappa\rangle}) \big) \bigg)        &                           \\
		      \end{cases}$
\end{enumerate}

Given the SMT encodings, we have all the ingredients necessary for assembling the bounded model checker for UCS, which we shall describe in the consequent section.

\section{Tool for verification of {\Lstar} properties}\label{ssec:verif}
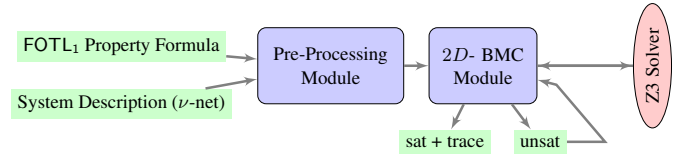
\begin{figure}[!ht]
    \centering
    \scalebox{0.8}{\tikzstyle{aldecision} = [diamond, draw, fill=blue!20,
text width=4.5em, text badly centered, node distance=2.5cm, inner sep=0pt]
\tikzstyle{alblock} = [rectangle, draw, fill=blue!20,
text width=5em, text centered, rounded corners, minimum height=4em]
\tikzstyle{line} = [draw, very thick, color=black!50, -latex']
\tikzstyle{allibrary} = [draw, ellipse,fill=red!20, node distance=2.5cm,
minimum height=2em]
\begin{tikzpicture}[node distance = 5em, auto, fit label/.style={yshift={(height("#1")+4pt)/2},
				inner ysep={(height("#1")+8pt)/2}, label={[anchor=north,font=\itshape]north:#1}}]

	\node [ node distance=2em, xshift=-3em,fill=green!20] (property) {{\Lstar} Property Formula};
	\node [ node distance=2em, below of=property, yshift=-1em, fill=green!20] (sysdesc) {System Description ($\nu$-net)};
	\node [alblock, node distance=7em, text width=7em,right of=property,yshift=-1em, xshift=4em] (ppm) {Pre-Processing\\ Module};

	\node [alblock, node distance=7em, right of=ppm, xshift=1em, minimum width=4em] (tool) {$2D$- BMC Module};

	\node [allibrary, right of= tool,xshift=1em,rotate=90](z3){Z3 Solver};

	\node [ node distance=2em, below of=tool,yshift=-2em,,xshift=-2em, fill=green!20] (sat) {sat + trace};
	\node [ node distance=2em, right of=sat,  xshift=3em,fill=green!20] (unsat) {unsat};

	\path [line] (sysdesc) -- (ppm);
	\path [line] (property) -- (ppm);

	\path [line] (ppm) -- (tool);

	\path [line] (z3) -- (tool);
	\path [line] (tool) -- (z3);
	\path [line] (tool) -- (sat);
	\path [line] (tool) -- (unsat);
	\path [line] (unsat) --([xshift=2em]unsat.east) -- ([xshift=1em,yshift=-1em]tool.east) -- (tool);

\end{tikzpicture}}
    \caption{UCSChecker  architecture}
    \label{fig:dc3arch}
\end{figure}

Using the SMT encoding, we demonstrate the verification of {\Lstar} properties using the bounded model checker, UCSChecker~\cite{ucsChecker}. Its architecture is in Fig.~\ref{fig:dc3arch}. UCSChecker has two primary inputs- system description in standard PNML format and the property to be tested expressed in {\Lstar} and the bound up to which the verification shall be performed. First, we validate and parse the  $\nu$-nets and {\Lstar} properties against their grammar  using ANTLR~\cite{ANTLRParrF11} and encode them into a formula.  Second, we perform bounded model checking, by bounding both the number of tokens and the time instance, which is called $2D$-BMC. The $2D$-BMC algorithm is introduced in~\cite{arxiv2dbmcv2}. We make use of the SAT/SMT Solver Z3~\cite{MouraB08}, to solve the encoded formula and give us a result of unsatisfiable, or satisfiable with a counterexample trace. If unsatisfiable, the tool can increment the bound and look further, until the external termination bound is hit.

While parsing the $\nu$-net, we validate it against the grammar for it. This contains both the lexer rules, which are used to build tokens and the parser rules, to validate the PNML file based on the generated parse tree. For instance, consider the following rule for validating the elements.
An element may be either an open tag followed by a close tag or an empty tag, each containing its name and possibly containing a set of attributes. There may be open tags for exactly one of the predefined keywords such as place, transition etc. Notice that it is not possible to define just a close tag using these set of rules. If such an input exists, our tool parses the PNML file and invalidates it by throwing a suitable error.
\begin{lstlisting}
header: '<?' Name attribute* '?>';
element:
       '<' Name attribute* '>' (
        place
        | transition
        | arc
        | element
        | textTagDigit
        | textTag
        | TEXT
        )* '<' '/' Name '>'
        | '<' Name attribute* '/>';
\end{lstlisting}
Similarly, we validate the {\Lstar} specifications as well. The implementation details of of the pre-processing module in UCSChecker and grammar used for validation are in Appendix.~\ref{lexParseRules}.

We verify the {\Lstar} properties of the APS case study and report them here. Since this is a first of its kind, there are no similar tools to compare the results of the model checker. However, we can see from the liveness windows that the counterexample or UNSAT obtained from the tool are justified.

\begin{figure}
    \centering
    \scalebox{0.7}{\input{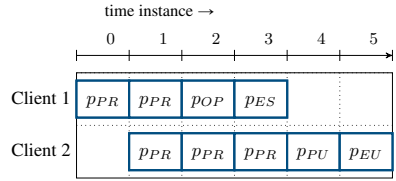}}
    \caption{Live window corresponding to the satisfaction of formula $\psi_1$}
    \label{fig:formula1}
\end{figure}

Given a snapshot of the running example Fig.~\ref{fig:formula1}, where there are two clients, consider the satisfaction of formula $\psi_1$, at an arbitrary bound $5$. Recall the formulas from Section~\ref{sec:mlogic}.

\begin{align*}
    \psi_1 & =\mathbf{G}_s(\forall x) \Big( parking\_requested(x) \Rightarrow               \\
           & \qquad\mathbf{F}_c~ (exit\_successfully(x)~\lor~exit\_unsuccessfully(x)) \Big)
\end{align*}

This can be paraphrased as follows:

$\psi_1 =\mathbf{G}_s(\forall x) \Big( PR(x) \Rightarrow \mathbf{F}_c~ (ES(x)~\lor~EU(x)) \Big)$

\noindent It can be observed that all the clients are in state $PR$ initially. Eventually, all the clients in the system are in either state $ES$ or $EU$. Here, $\psi_1$ holds true.

\begin{figure}
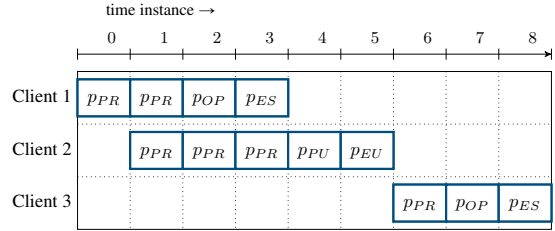

    \centering
    \scalebox{0.7}{\begin{ganttchart}[
        vgrid,hgrid,
        progress label text=\relax,
        x unit=10mm,
        title height=1,
        foobar top shift=.15,
        foobar height=.70
    ]{0}{8}
    \textganttbar{Client 1}{$p_{PR}$}{0}{0}  \textganttbar{}{$p_{PR}$}{1}{1} \textganttbar{}{$p_{OP}$}{2}{2} \textganttbar{}{$p_{ES}$}{3}{3} \\
    \textganttbar{Client 2}{$p_{PR}$}{1}{1}  \textganttbar{}{$p_{PR}$}{2}{2}  \textganttbar{}{$p_{PR}$}{3}{3} \textganttbar{}{$p_{PU}$}{4}{4} \textganttbar{}{$p_{EU}$}{5}{5}\\

    \textganttbar{Client 3}{$p_{PR}$}{6}{6}  \textganttbar{}{$p_{OP}$}{7}{7} \textganttbar{}{$p_{ES}$}{8}{8}
    \begin{scope}[yshift=0.33cm]
        \draw[->] (0,0) -- (9,0);
        \draw ([yshift=-3pt]0,0) -- ++([yshift=6pt]0,0) node[xshift=45pt, yshift=15pt, above] {time instance $\rightarrow$};
        \draw ([yshift=-3pt]1,0) -- ++([yshift=6pt]0,0) node[xshift=-10pt, above] {$0$};
        \draw ([yshift=-3pt]2,0) -- ++([yshift=6pt]0,0) node[xshift=-10pt, above] {$1$};
        \draw ([yshift=-3pt]3,0) -- ++([yshift=6pt]0,0) node[xshift=-10pt, above] {$2$};
        \draw ([yshift=-3pt]4,0) -- ++([yshift=6pt]0,0) node[xshift=-10pt, above] {$3$};
        \draw ([yshift=-3pt]5,0) -- ++([yshift=6pt]0,0) node[xshift=-10pt, above] {$4$};
        \draw ([yshift=-3pt]6,0) -- ++([yshift=6pt]0,0) node[xshift=-10pt, above] {$5$};
        \draw ([yshift=-3pt]7,0) -- ++([yshift=6pt]0,0) node[xshift=-10pt, above] {$6$};
        \draw ([yshift=-3pt]8,0) -- ++([yshift=6pt]0,0) node[xshift=-10pt, above] {$7$};
        \draw ([yshift=-3pt]9,0) -- ++([yshift=6pt]0,0) node[xshift=-10pt, above] {$8$};
    \end{scope}
\end{ganttchart}}
    \caption{Snapshot of APS with three clients}
    \label{fig:formula2}
\end{figure}

Given a snapshot of the running example Fig.~\ref{fig:formula2}, where there are three clients, consider the satisfaction of formula $\psi_2$, at an arbitrary bound $5$.

\begin{align*}
    \psi_2 & =\mathbf{G}_s(\forall x) \Big(occupy\_parking\_lot(x)\Rightarrow \mathbf{F}_c(exit\_successfully(x))\Big)
\end{align*}

This can be paraphrased as follows:

$\psi_2 =\mathbf{G}_s(\forall x) \Big( OP(x) \Rightarrow \mathbf{F}_c~ES(x) \Big)$

\noindent It can be observed that clients $1$ and $2$ are at $OP$ and eventually go to state $ES$. Hence, $\psi_2$ holds true.

Consider another formula:
\begin{align*}
    \psi_{3} & =\mathbf{G}_s(\exists x) \big(parking\_requested(x) \land \mathbf{F}_c (exit\_unsuccessfully(x))\big)
\end{align*}

This can be paraphrased as follows:

$\psi_3 =\mathbf{G}_s(\exists x) \big(PR(x) \land \mathbf{F}_c (EU(x))\big)$

In the same Fig.~\ref{fig:formula2}, client $2$ is in state $PR$ at instance $1$ through $3$ and eventually in state $EU$ at instance $5$. Hence, formula $\psi_3$ is satisfied at instance $5$.

Consider the following formula:
\begin{align*}
    \psi_4 & = \mathbf{G}_s(\exists x) \bigg( parking\_requested(x)~\land                                      \\
           & \qquad \mathbf{F}_c \big( parking\_unavailable(x)~\mathbf{U}_c~exit\_unsuccessfully(x)\big)\bigg)
\end{align*}

This can be paraphrased as follows:

$\psi_4=\mathbf{G}_s(\exists x) \bigg( PR(x)~\land
    \mathbf{F}_c \big( PU(x)~\mathbf{U}_c~EU(x)\big)\bigg)$

In Fig.~\ref{fig:formula2}, formula $\psi_4$ is also satisfied at instance $5$.
The bounded model checker UCSChecker is hosted on the public domain such that the results are verifiable. It can also verify other $\nu$-nets apart from UCS and their properties expressed in {\Lstar}.

\subsection{Case Study: Travel Agency}\label{ssec:casestudytravel}

\begin{figure}[h]
    \begin{center}
        \scalebox{0.8}{\begin{tikzpicture}[scale=0.6]
    \centering
    \node[transition,fill=black,label=below:$t_{src}$,inner
        sep=0pt,minimum height=.40cm, minimum width=.5cm] (tsrc) at (0,1){} ;

    \node[place,draw=red!80,minimum height=.4cm,minimum width=.9cm,label=left:{\small $p_{il}$}] (pil) at (-4,-1) {$\{1\}$} edge[pre] node[auto] {$\nu$} (tsrc);

    \node[place,draw=blue!80,minimum height=.4cm,minimum width=.9cm,label=left:{\small $p_{sr}$}] (psr) at (-12,-1) {$\{0\}$};

    \node[transition,fill=black,label=below:$t_{st}$,inner
        sep=0pt,minimum height=.40cm, minimum width=.5cm] (tst) at (0,-1){} edge[pre] node[auto] {c} (pil);
    \node[place,draw=red!80,minimum height=.4cm,minimum width=.9cm,label=right:{\small $p_{ss}$}] (pss) at (2,-3) {} edge[pre] node[auto] {c} (tst);
    \node[transition,fill=black,label=below:$t_{lf}$,inner
        sep=0pt,minimum height=.40cm, minimum width=.5cm] (tlf) at (0,-4){} edge[pre] node[auto] {c} (pss);
    \node[place,draw=red!80,minimum height=.4cm,minimum width=.9cm,label=below:{\small $p_{bf}$}] (pbf) at (-2,-4) {} edge[pre] node[auto] {c} (tlf);
    \node[transition,fill=black,label=left:$t_{vf}$,inner
        sep=0pt,minimum height=.40cm, minimum width=.5cm] (tvf) at (-4,-4){} edge[pre] node[auto] {c} (pbf) edge[pre] node[auto,pos=0.2] {c} (pil) edge[pre] node[auto] {s} (psr);

    \node[transition,fill=black,label=below:$t_{lb}$,inner
        sep=0pt,minimum height=.40cm, minimum width=.5cm] (tlb) at (0,-7){} edge[pre] node[auto] {c} (pss);
    \node[place,draw=red!80,minimum height=.4cm,minimum width=.9cm,label=below:{\small $p_{bb}$}] (pbb) at (-2,-7) {} edge[pre] node[auto] {c} (tlb);
    \node[transition,fill=black,label=left:$t_{vb}$,inner
        sep=0pt,minimum height=.40cm, minimum width=.5cm] (tvb) at (-8,-7){} edge[pre] node[auto,pos=0.2] {c} (pbb) edge[pre] node[auto,pos=0.1] {c} (pil) edge[pre] node[auto] {s} (psr);

    \node[transition,fill=black,label=below:$t_{lt}$,inner
        sep=0pt,minimum height=.40cm, minimum width=.5cm] (tlt) at (0,-10){} edge[pre] node[auto] {c} (pss);
    \node[place,draw=red!80,minimum height=.4cm,minimum width=.9cm,label=below:{\small $p_{bt}$}] (pbt) at (-2,-10) {} edge[pre] node[auto] {c} (tlt);
    \node[transition,fill=black,label=left:$t_{vt}$,inner
        sep=0pt,minimum height=.40cm, minimum width=.5cm] (tvt) at (-12,-10){} edge[pre] node[auto,pos=0.1] {c} (pbt) edge[pre] node[auto,pos=0.1] {c} (pil) edge[pre] node[auto,pos=0.2] {s} (psr);

    \node[place,draw=blue!80,minimum height=.4cm,minimum width=.9cm,label=left:{\small $p_{sa}$}] (psa) at (-10,-14) {}
    edge[pre] node[auto,pos=0.1] {s} (tvf)
    edge[pre] node[auto,pos=0.1] {s} (tvb)
    edge[pre] node[auto,pos=0.1] {s} (tvt);

    \node[place,draw=red!80,minimum height=.4cm,minimum width=.9cm,label=left:{\small $p_{cp}$}] (pcp) at (-6,-14) {}
    edge[pre] node[auto,pos=0.1] {c} (tvf)
    edge[pre] node[auto,pos=0.1] {c} (tvb)
    edge[pre] node[auto,pos=0.1] {c} (tvt);

    \node[transition,fill=black,label=below:$t_{vp}$,inner
        sep=0pt,minimum height=.40cm, minimum width=.5cm] (tvp) at (-8,-16){}
    edge[pre] node[auto,pos=0.5] {s} (psa)
    edge[pre] node[auto,pos=0.5] {c} (pcp);

    \node[place,draw=blue!80,minimum height=.4cm,minimum width=.9cm,label=below:{\small $p_{rp}$}] (prp) at (-10,-18) {}
    edge[pre] node[auto,pos=0.5] {s} (tvp);

    \node[place,draw=red!80,minimum height=.4cm,minimum width=.9cm,label=below:{\small $p_{es}$}] (pes) at (-6,-18) {}
    edge[pre] node[auto,pos=0.5] {c} (tvp);

    \node[transition,fill=black,label=below:$t_{exit}$,inner
        sep=0pt,minimum height=.40cm, minimum width=.5cm] (texit) at (-2,-18){}
    edge[pre] node[auto] {d} (pes);

    \node[transition,fill=black,label=below:$t_{udb}$,inner
        sep=0pt,minimum height=.40cm, minimum width=.5cm] (tudb) at (-16,-18){}
    edge[pre] node[auto] {s} (prp)
    edge[->] node[auto] {s} (psr);

\end{tikzpicture}}
        \caption{The $\nu$-net modeling a travel agency system}
        \label{fig:travelagencyv0}
    \end{center}
\end{figure}
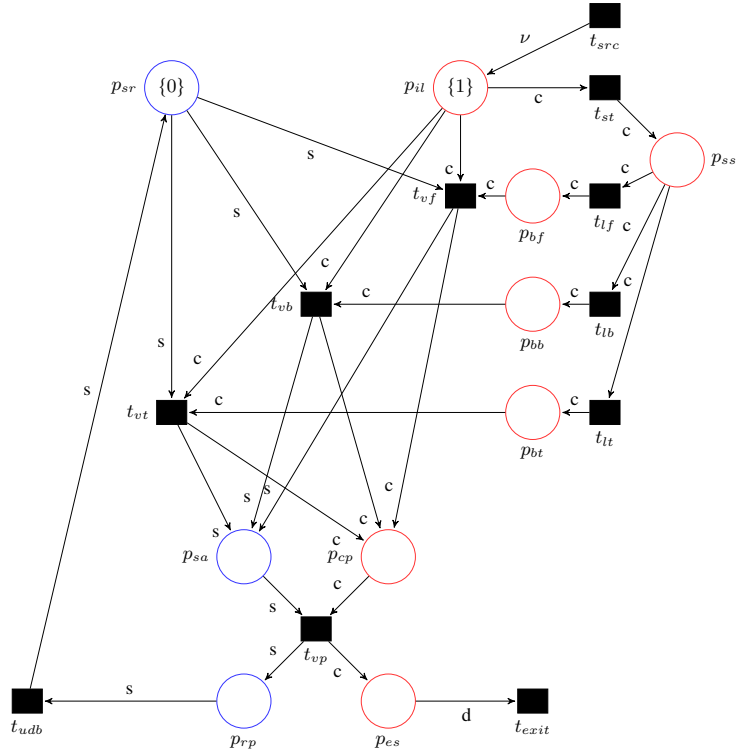
A travel agency is represented in the $\nu$-net in Fig.~\ref{fig:travelagencyv0}. The source transition $t_{src}$ adds new tokens (representing clients) into the system, to the place $p_{il}$ which denotes the clients which have entered the system. The transition $t_{st}$ denotes the \emph{search trip} activity. After the clients have \emph{searched succesfully}, they enter the place $p_{ss}$. From here, they have options to choose the mode of the trip, either the activity \emph{lookup flight} or \emph{lookup bus} or \emph{lookup train} as represented by the transitions $t_{lf}$, $t_{lb}$ and $t_{lt}$. On choosing either of these, the clients move to the respective \emph{book flight}, \emph{book bus}, or \emph{book train} represented by places $p_{bf}$, $p_{bb}$ and $p_{bt}$ respectively. The transitions $t_{vf}$, $t_{vb}$ and $t_{vt}$ validate the flight, bus or train journeys respectively using the information provided by the server from place $p_{sr}$ when the server gives the response. On validating the journey information, the server gathers the client information for \emph{server approval} in place $p_{sa}$ and the \emph{client shares payment information} in place $p_{cp}$. Following this, they \emph{verify payment} as given in transition $t_{vp}$ and the server receives the response in place $p_{rp}$ and accordingly updates its database using transition $t_{udb}$ and gets ready for the next client interaction, in place $p_{sr}$. The client exits the system via place $p_{es}$ and gets terminated using transition $t_{exit}$.
We list some of the interesting properties of this system:
\begin{enumerate}
    \item If a client has entered the system, then eventually it will exit the system. This can be expressed as $\Phi_1$=$\mathbf{G}_s(\forall x) \big( p_{il}(x) \Rightarrow \mathbf{F}_c~p_{es}(x) \big)$.
    \item  If a client has entered the system, then it will not be in the place $p_{il}$ after exiting the system. This can be expressed as $\Phi_2$=$\mathbf{G}_s(\forall x) \big( p_{es}(x) \Rightarrow \neg p_{il}(x) \big)$.
    \item  If a client has entered the system and it has chosen the mode of journey as flight, then eventually it will be in the place $p_{cp}$, where they share their payment information. This can be expressed as $\Phi_3$=$\mathbf{G}_s(\forall x) \big( p_{il}(x) \land p_{bf}(x)\Rightarrow \mathbf{F}_c~p_{cp}(x) \big)$.
    \item  If a client has entered the system and it has chosen the mode of journey as flight, and its journey is not yet validated by the server, then eventually it will be in the place $p_{cp}$, where they share their payment information to the server. This can be expressed as $\Phi_4$=$\mathbf{G}_s(\forall x) \big( p_{il}(x) \land p_{bf}(x) \land \neg p_{cp}(x)\Rightarrow \mathbf{F}_c~p_{cp}(x) \big)$.

    \item  If a client has entered the system, then eventually it will be in the place $p_{cp}$, after which it gets server approval. This can be expressed as $\Phi_5$=$\mathbf{G}_s(\forall x) \big( p_{il}(x) \Rightarrow \mathbf{F}_c~p_{cp}(x) \big)$.
    \item If a client's payment has been validated, then it will not be in the place $p_{cp}$. This can be expressed as $\Phi_6$=$\mathbf{G}_s(\forall x) \big( p_{es}(x) \Rightarrow \neg p_{cp}(x) \big)$.
    \item If a client is searching for a journey it can only choose either of the three flight, train or bus journeys. $\Phi_7$=$\mathbf{G}_s(\forall x) \big( p_{ss}(x) \Rightarrow F_c( p_{bf}(x) \lor p_{bt}(x) \lor p_{bb}(x) \big)$.
\end{enumerate}

\subsection{Experimental results}\label{ssec:experimentsfotl}
We test the properties of the case studies written using {\Lstar} using our open source tool UCSChecker~\cite{ucsChecker}. The tool provides the detailed counterexample trace when the property is violated. We assume a bound of $5$ for the below experiments.  In order to verify properties of other $\nu$-nets, one needs to create the model using a PN editor and then specify the properties in {\Lstar}.
\begin{table}[h!]
    \centering
    \caption{Experiments on APS case study}
    \label{tab:simple-table}
    \begin{tabular}{|c|c|c|c|}
        \hline
        Property ID & Maximum Nesting Depth of temporal operators & Number of clauses & time (s) \\
        \hline
        $\Psi_1$    & 2                                           & 3                 & 0.32     \\

        \hline
        $\Psi_2$    & 2                                           & 2                 & 0.4      \\
        \hline
        $\Psi_3$    & 2                                           & 2                 & 0.3      \\
        \hline
        $\Psi_4$    & 2                                           & 3                 & 0.38     \\
        \hline
    \end{tabular}
\end{table}

\begin{table}[h!]
    \centering
    \caption{Experiments on travel agency case study}
    \label{tab:exptravelagency}
    \begin{tabular}{|c|c|c|c|}
        \hline
        Property ID & Maximum Nesting Depth of temporal operators & Number of clauses & time (s) \\
        \hline
        $\Phi_1$    & 2                                           & 2                 & 0.027    \\

        \hline
        $\Phi_2$    & 1                                           & 2                 & 0.028    \\
        \hline
        $\Phi_3$    & 2                                           & 3                 & 0.032    \\
        \hline
        $\Phi_4$    & 2                                           & 4                 & 0.028    \\
        \hline
        $\Phi_5$    & 2                                           & 2                 & 0.025    \\
        \hline
        $\Phi_6$    & 1                                           & 2                 & 0.025    \\
        \hline
        $\Phi_7$    & 2                                           & 4                 & 0.028    \\
        \hline
    \end{tabular}
\end{table}

\subsection{Limitations of the tool}\label{ssec:limitationsfotltool}
The current toolchain depends on a PNML Editor, Wolfgang~\cite{PNWolfgang}, which we use to draw the net and mention the arc labels. However, identifiable tokens are not supported by the Wolfgang editor, as it is primarily a Petri net and Colored Petri net simulation tool. Hence, we have to handcraft the $\nu$-net arcs, which is not sustainable when the number of places and transitions are huge.
While there are several tools to model other types of PNs, to the best of our knowledge, there are none that provide a graphical representation of $\nu$-nets. We consider this as a future extension to this work. While the logic {\Lstar} allows us to reason about distinguishable clients, we cannot describe properties of each individual client. In future, we would like to explore different logics to do the same.
\section{Conclusion and Future Work}\label{sec:concl}

In this work, we propose a one-variable fragment of $MFOTL$, called {\Lstar} for unbounded client-server systems modeled using $\nu$-nets and build a bounded model checking tool that can check properties of {\Lstar} using SMT encoding. In~\cite{HKKWZ}, it has been shown that the one variable fragment of MFOTL fragment is also $EXPSPACE$-complete. It remains to show if {\Lstar} is also $EXPSPACE$-complete.

Building verification tools for unbounded client-server systems is a non-trivial engineering pursuit. The BMC implementation using {\Lstar} specifications will join the arsenal of tools such as KREACH~\cite{DixonL20}, Petrinizer~\cite{EsparzaLMMN14}, QCOVER~\cite{BlondinFHH16}, ICOVER~\cite{GeffroyLS18} for nets.
We can also have multiple types of clients in the unbounded client-server systems, which may model a richer set of systems.

\bibliographystyle{splncs04}
\bibliography{references}
\appendix
\section{Satisfiability relations for {\Lstar}}\label{sec:unboundedsem}
We describe the satisfiability relations for the logic.
Let $M=(\nu,V,\xi)$ be a valid model. The satisfiability relations $\models$  and $\models_x$ can be defined, via induction over the structure of $\psi\in \Psi$,  and $\alpha\in \Delta$, respectively, as follows:
\begin{enumerate}
    \item $M,i\models q$ iff $q \in \nu_i$.

    \item $M,i\models \lnot q$ iff $q \not\in \nu_i$.

    \item $M,i \models (\exists x)\alpha$ iff $\exists a\in CN$, $a \in V_i$ and $M,[x\mapsto a],i\models_x \alpha$.

    \item $M,i \models (\forall x)\alpha$ iff $\forall a\in CN$, if $a \in V_i$ then $M,[x\mapsto a],i\models_x \alpha$.

    \item $M,i\models \psi_1\lor\psi_2$ iff $M,i\models \psi_1$ or $M,i\models \psi_2$.

    \item $M,i\models \psi_1\land\psi_2$ iff $M,i\models \psi_1$ and $M,i\models \psi_2$.

    \item $M,i\models \mathbf{X_s}\psi$ iff $M,i+1\models \psi$.

    \item $M,i\models \mathbf{F_s}\psi$ iff $\exists j \ge i$, $M,j\models \psi$.

    \item $M,i\models \mathbf{G_s} \psi$ iff $\forall j \ge i$, $M,j\models \psi$.

    \item $M,i\models \psi_1\mathbf{U_s}\psi_2$ iff $\exists j \ge i$, $M,j\models \psi_2$ and $\forall j':i\le j' < j: M,j' \models \psi_1$.

    \item[]

    \item $M,[x\mapsto a],i\models_x p(x)$ iff $\xi_i(a,p)=\top$.

    \item $M,[x\mapsto a],i\models_x \lnot \alpha$ iff $M,[x\mapsto a],i\not\models_x \alpha$.

    \item $M,[x\mapsto a],i\models_x \alpha\lor \beta$ iff $M,[x\mapsto a],i\models_x\alpha$ or $M,[x\mapsto a],i\models_x\beta$.

    \item $M,[x\mapsto a],i\models_x \alpha\land \beta$ iff $M,[x\mapsto a],i\models_x\alpha$ and $M,[x\mapsto a],i\models_x\beta$.

    \item $M,[x\mapsto a],i\models_x \mathbf{X}_c\alpha$ iff $M,[x\mapsto a],i+1\models_x \alpha$ and $a\in V_{i+1}$.

    \item $M,[x\mapsto a],i\models_x \mathbf{F}_c\alpha$ iff $\exists j\ge i$, $a\in V_j$, $M,[x\mapsto a],j\models_x \alpha$ .

    \item $M,[x\mapsto a],i\models_x \mathbf{G}_c\alpha$ iff $\forall j\ge i$, $a\in V_j$, $M,[x\mapsto a],j\models_x \alpha$.

    \item $M,[x\mapsto a],i\models_x \alpha\mathbf{U}_c\beta$ iff $\exists j \ge i$, $M,[x\mapsto a],j\models_x \beta$ and $\forall j':i\le j'  < j: M,j' \models \alpha$ and $a\in V_j$.

\end{enumerate}

We make use of the satisfiability relations described above, to check the model M with respect to a formula at an instance $i$. This is necessary for arriving at the bounded semantics for {\Lstar}.

\section{Bounded Semantics of {\Lstar}}\label{sec:boundedsemboth}
We describe the bounded semantics of {\Lstar} in two sections, based on the existence of a loop in the behaviour of the net and without a loop.
\subsection{Bounded Semantics Without Loop}\label{ssec:boundedloopfree}
First, we describe the bounded semantics without loop as shown in Fig.~\ref{fig:loopfreepath} where $0\leq i \leq \lambda$, where $i$ is the current instance on the bounded path. We introduce $\models^{k}_x$ as an extension of the satisfaction relation $\models^{k}$ with respect to the free variable $x$, where $x$ is used to denote the client names.

\begin{enumerate}
    \item $M,i\models^{k} q$ iff $q \in \nu_i$.

    \item $M,i\models^{k} \lnot q$ iff $q \not\in \nu_i$.

    \item $M,i \models^{k} (\exists x)\alpha$ iff $\exists a\in CN$, $a \in V_i$ and $M,[x\mapsto a],i\models^{k}_x \alpha$.

    \item[] The above formula is satisfied when there is at least one live client such that $\alpha$ is satisfied in the model at instance $i$ for the particular live client $a$.

    \item $M,i \models^{k} (\forall x)\alpha$ iff $\forall a\in CN$, if $a \in V_i$ then $M,[x\mapsto a],i\models^{k}_x \alpha$.

    \item[] The above formula is satisfied when the $\alpha$ is satisfied in the model at instance $i$ for all live clients in the set $CN$.

    \item $M,i\models^{k} \psi_1\lor\psi_2$ iff $M,i\models^{k} \psi_1$ or $M,i\models^{k} \psi_2$.

    \item $M,i\models^{k} \psi_1\land\psi_2$ iff $M,i\models^{k} \psi_1$ and $M,i\models^{k} \psi_2$.

    \item $M,i\models^{k} \mathbf{X_s}\psi$ iff $
              \begin{cases}
                  M,i+1\models^{k} \psi                & \text{if }(i<\lambda) \\
                  M,i\not \models^{k}\mathbf{X_s} \psi & \text{otherwise }
              \end{cases}$

    \item[] When instance $i$ is less than the bound $\lambda$, the formula $\psi$ is evaluated at instance $i+1$ else, the formula is unsatisfiable.

    \item $M,i\models^{k} \mathbf{F_s}\psi$ iff $\exists j: i \le j \le\lambda$ , $M,j\models^{k} \psi$.

    \item[] This formula is satisfiable if there is some instance $j$ such that $i \le j \le\lambda$ at which the property $\psi$ holds.

    \item $M,i\not\models^{k} \mathbf{G_s} \psi$

    \item[] In the absence of a loop in the bounded run, the above formula is always unsatisfiable, since global server behaviour cannot be evaluated without a loop.

    \item $M,i\models^{k} \psi_1\mathbf{U_s}\psi_2$ iff $\exists j: i \le j \le\lambda$, $M,j\models^{k} \psi_2$ and $\forall j': i \le j' < j: M,j' \models^{k} \psi_1$.

    \item[] This formula is satisfied when formula $\psi_2$ is satisfied at some instance $j$  and for all instances less than $j$, formula $\psi_1$ is satisfied.

    \item $M,[x\mapsto a],i\models^{k}_x p(x)$ iff $\xi_i(a,p)=\top$.

    \item[] This formula is satisfied if the property $p$ is satisfied by live agent $a$ at instance $i$. Here, the free variable $x$ is substituted by the client name $a$.
    \item $M,[x\mapsto a],i\models^{k}_x \lnot \alpha$ iff $M,[x\mapsto a],i\not\models^{k}_x \alpha$.

    \item $M,[x\mapsto a],i\models^{k}_x \alpha\lor \beta$ iff $M,[x\mapsto a],i\models^{k}_x\alpha$ or $M,[x\mapsto a],i\models^{k}_x\beta$.

    \item $M,[x\mapsto a],i\models^{k}_x \alpha\land \beta$ iff $M,[x\mapsto a],i\models^{k}_x\alpha$ and $M,[x\mapsto a],i\models^{k}_x\beta$.

    \item $M,[x\mapsto a],i\models^{k}_x \mathbf{X}_c\alpha$ iff $
              \begin{cases}
                  M,[x\mapsto a],i+1\models^{k}_x \alpha                 & \text{if } (i<\lambda \text{ and } a \in V_{i+1}) \\
                  M,[x\mapsto a],i\not \models^{k}_x \mathbf{X}_c \alpha & \mbox{otherwise }
              \end{cases}$

    \item[] If the instance $i$ is less than the bound $\lambda$ and the client is a live client at instance $i+1$, then the formula is evaluated at the instance $i+1$ for the live client $a$; else the original formula is not satisfied.

    \item $M,[x\mapsto a],i\models^{k}_x \mathbf{F}_c\alpha$ iff $\exists j: i \le j \le\lambda$, $a\in V_j$ and $M,[x\mapsto a],j\models^{k}_x \alpha$.

    \item[] This formula is satisfiable if there is some instance $j$ such that $i \le j \le\lambda$ at which the property $\alpha$ is satisfied for the client $a$ and the client $a$ is a live client at instance $j$ .

    \item $M,[x\mapsto a],i \models^{k}_x \mathbf{G}_c\alpha \text{ iff }\\
              \begin{cases}
                  M,[x\mapsto a],j \not \models^{k}_x \alpha                           & \text{if }(\forall j> i \text{ and } a\in V_j) \\
                  \forall j':i\le j'<j, M,[x\mapsto a],j' \models^{k}_x \alpha         & \text{otherwise }                              \\
                  \text{where } j>i \text{ be the least instance where }a \not \in V_j &
              \end{cases}$

    \item[] For all instances $j$ such that $j>i$ if client $a$ is live at instance $j$, the formula is not satisfiable at $j$.
        Let $j$ be the least instance where client $a$ is not live. For all instances $j'$ such that $i\le j'<j$, formula $\alpha$ is evaluated at $j'$.

    \item $M,[x\mapsto a],i\models^{k}_x \alpha\mathbf{U}_c\beta$ iff $\exists j \ge i$, $a\in V_j$,  $M,[x\mapsto a],j\models^{k}_x \beta$, and $\forall j': i \le j'  < j$,  $M,[x\mapsto a],j' \models^{k} \alpha$.

    \item[] This is satisfied when $\beta$ is satisfied at some instance $j$ and client $a$ is a live client at $j$; and for all instances less than $j$, client formula $\alpha$ is satisfied.

\end{enumerate}
\subsection{Bounded Semantics with Loop}\label{ssec:boundedLoop}
\begin{enumerate}

    \item[7.] $M,i\models^{k} \mathbf{X_s}\psi$ iff $
            \begin{cases}
                M,i+1\models^{k} \psi & \text{ if } (i <\lambda) \\
                M,l\models^{k} \psi   & \text{otherwise }
            \end{cases}$

    \item[] When instance $i$ is less than the bound $\lambda$, the formula $\psi$ is evaluated at instance $i+1$ else, $\psi$ is evaluated at instance $l$, which is the next instance of $\lambda$.

    \item[8.] $M,i\models^{k} \mathbf{F_s}\psi$ iff $\exists j: min(l,i)\le j \le \lambda$, $M,j\models^{k} \psi$.

    \item[] This formula is satisfiable if there is some instance $j$ such that $min(l,i) \le j \le\lambda$, where the formula $\psi$ is satisfied.

    \item[9.] $M,i\models^{k} \mathbf{G_s} \psi$ iff $\forall j: min(l,i)\le j \le \lambda$, $M,j\models^{k} \psi$.

    \item[] This formula is satisfiable if for all instances $j$ such that $min(l,i) \le j \le\lambda$, the formula $\psi$ is satisfied in each instance.

    \item[10.] $M,i\models^{k} \psi_1\mathbf{U_s}\psi_2 \text{ iff }
            \begin{cases}
                \exists j:i \le j \le \lambda, M,j\models^{k} \psi_2 \text{ and } & \text{ if }(i\le l) \\
                \forall j': i \le j' < j: M,j' \models^{k} \psi_1                 &                     \\
                \exists j: i \le j \le \lambda, M,j\models^{k}\psi_2 \text{ and } & \text{ if } (i> l)  \\
                \forall j': i \le j' <j: M,j' \models^{k} \psi_1                  &                     \\
                \hspace{10em}\text{or}                                            &                     \\
                \exists j: l \le j < i, M,j\models^{k}\psi_2 \text{ and }         &                     \\
                \forall j': l \le j' < j: M,j' \models^{k} \psi_1
            \end{cases}$

    \item[] Consider the two cases:
        \begin{itemize}
            \item If $(i\le l)$, the current instance $i$ is less than or equal to the loop instance $l$, this formula is satisfied when formula $\psi_2$ is satisfied at some instance $j$  such that $i \le j \le \lambda$ and for all instances between $i$ and $j$, formula $\psi_1$ is satisfied.
            \item If $(i> l)$, the formula may be satisfied in either of the two intervals between $i$ to $\lambda$ or between $l$ to $i$. Hence, the formula is satisfied if either of the following is satisfied: formula $\psi_2$ is satisfied at some instance $j$  such that $i \le j \le \lambda$ and for all instances less than $j$, formula $\psi_1$ is satisfied or, formula $\psi_2$ is satisfied at some instance $j$  such that $l \le j < i$ and for all instances between $l$ and $j$, formula $\psi_1$ is satisfied.
        \end{itemize}

    \item[15.] $M,[x\mapsto a],i\models^{k}_x \mathbf{X}_c\alpha \text{ iff }
            \begin{cases}
                M,[x\mapsto a],i+1\models^{k}_x \alpha                & \text { if } i <\lambda \text{ and }a \in V_{i+1} \\
                M,[x\mapsto a],l\models^{k}_x \alpha                  & \text{ if } i = \lambda \text{ and }a \in V_{l}   \\
                M,[x\mapsto a],i\not\models^{k}_x \mathbf{X}_c \alpha & \text{ otherwise }
            \end{cases}$
    \item[] When current instance $i$ is less than the bound $\lambda$, and the client $a$ is a live agent at the next instance $i+1$, the formula is evaluated at instance $i+1$ else, in the current instance if the bound $\lambda$ and the client is live at instance $l$, then the formula is evaluated at instance $l$, which is the next instance of $\lambda$. Otherwise, the original formula does not hold.

    \item[16.] $M,[x\mapsto a],i\models^{k}_x \mathbf{F}_c\alpha$ iff $\exists j: min(l,i) \le j \le\lambda$ , $a\in V_j$ , $M,[x\mapsto a],j\models^{k}_x \alpha$.

    \item[] This formula is satisfiable if there is some instance $j$ such that $min(l,i) \le j \le\lambda$, and the client $a$ is live at instance $j$, where the formula $\psi$ is satisfied.

    \item[17.] $M,[x\mapsto a],i\models^{k}_x \mathbf{G}_c\alpha$ iff $\forall j: min(l,i) \le j \le\lambda$ , $a\in V_j$ , $M,[x\mapsto a],j\models^{k}_x \alpha$.

    \item[18.] $M,[x\mapsto a],i\models^{k} \alpha\mathbf{U_c}\beta \text{ iff }\\
            \begin{cases}
                \exists j:i \le j \le \lambda,a\in V_j, M,[x\mapsto a],j\models^{k} \beta \text{ and } & \text{ if }(i\le l) \\
                \forall j': i \le j' < j: M,[x\mapsto a],j' \models^{k} \alpha                         &                     \\
                \exists j: i \le j \le \lambda,a\in V_j, M,[x\mapsto a],j\models^{k}\beta \text{ and } & \text{ if } (i> l)  \\
                \forall j': i \le j' <j: M,[x\mapsto a],j' \models^{k} \alpha                          &                     \\
                \hspace{10em}\text{or}                                                                 &                     \\
                \exists j: l \le j < i$,$a\in V_j, M,[x\mapsto a],j\models^{k}\beta \text{ and }       &                     \\
                \forall j': l \le j' < j, M, [x\mapsto a], j'\models^{k} \alpha.
            \end{cases}$

    \item[] Consider the two cases:
        \begin{itemize}
            \item If $(i\le l)$, the current instance $i$ is less than or equal to the loop instance $l$, and the client $a$ is live at instance $j$ then this formula is satisfied when formula $\psi_2$ is satisfied at some instance $j$  such that $i \le j \le \lambda$ and for all instances between $i$ and $j$, formula $\psi_1$ is satisfied.
            \item If $(i> l)$, the formula may be satisfied in either of the two intervals between $i$ to $\lambda$ or between $l$ to $i$. Hence, the formula is satisfied if either of the following is satisfied: formula $\psi_2$ is satisfied at some instance $j$  such that $i \le j \le \lambda$ and the client $a$ is live at instance $j$ and for all instances less than $j$, formula $\psi_1$ is satisfied or, formula $\psi_2$ is satisfied at some instance $j$ such that $l \le j < i$ and the client $a$ is live at instance $j$ and for all instances between $l$ and $j$, formula $\psi_1$ is satisfied.
        \end{itemize}
\end{enumerate}

\subsection{Representation of $\nu$-nets}\label{sec:smtnu}

In the context of verification of $\nu$-nets, we discuss the representation of nets such that they can be added as constraints in an SMT solver.

We outline the variables and data structures that are necessary to describe the true concurrent semantics. We have a finite set of transitions, $t_0,\ldots,t_m$ described in the net, their names are in the list $tNames$. We have a finite set of places, $p_0,\ldots,p_l$ their names are in the list $pNames$. Arcs can be of either of two types: where the source is a transition and the target is a place or the source is a place and the target is a transition. In case of the $\nu$-net representing the running example of UCS, depending on the mode, we also account for the label of the arcs and add an additional constraint, allowing only the tokens to move between places of the same type, i.e, tokens move from client (server) places to client (server) places. Additionally, we have a datastructure to keep track of the token identifiers, which is a list of integers. We allocate token $0$, to represent the server process, in our single server UCS. All other token identifiers belong to the client processes. The net formalism does not allow arcs between places and between transitions themselves. If they occur, the net description is erroneous, and we cannot move ahead with the unfolding.

We use the vector $iWTk$ to store the expressions with respect to $k$, to compute the incident weights of transitions. We use the vector $WTk$ to store the expressions with respect to $k$, to denote the change in weights for places.
We use two-dimensional matrices $Wt[m][l]$ and $iWt[m][l]$ to denote the net change in weights and the incidence weights (outgoing from places). The expressions for the same are stored in $WTVars$ and $iWTVars$ respectively.

We construct the two-dimensional weight matrix $Wt[row][col]$ of size $m \times l$ which contains the net weight of the arcs. Initially, all the matrix entries are initialized to zero.

For an arc from place $p_i$ to transition $t_j$ with the weight $w$, we have the matrix entry $Wt[i][j] = Wt[i][j] - w$.
For an arc from transition $t_j$ to place $p_i$ with the weight $w$, we have the matrix entry $Wt[i][j] = Wt[i][j]  + w$. Now, if there are incoming and outgoing arcs of the same weights, then the net weight $Wt[i][j]=0$. Notice that, by looking only at the Wt matrix, we may not distinguish between the case where there are no arcs to and from an element of the net. Hence, it is necessary to have a separate data structure for the same. We have a two-dimensional matrix $iWt[row][col]$ of size $m \times l$ which contains the incident weights to the transitions.
For an arc from place $p_i$ to transition $t_j$ with the weight $w$, we have the matrix entry $iWt[i][j] =Wt[i][j] - w$. The marking of the net consists of the set tokens at each place and represents the state of the net at any instance.
The initial marking of the net can be obtained from the net description and contains the number of tokens in each of the places $p_0,\ldots,p_l$ in the net. The subsequent markings may be constructed from the matrix $Wt$ and using the transition function of the net. The transition function $TF$ describes the behaviour of the net. The initial marking of the net is stored in $initial$. We introduce a method $printTFTruthTable$ that can aid to visualise the transition function $TF$ which is an expression describing the function. The variable $T[ti]$ denotes the $i$th transition being fired. Hence the expression $!T[ti]$ denotes that the $i$-th transition is not fired. For every pair of transitions and places, if there are no outgoing arc from the transition $t_i$ then the expression $preCond$ containing the precondition for firing of transitions is constructed as follows:

\begingroup
\allowdisplaybreaks
\begin{align}
    if (emptyOutTi)     & \{                                                  \\
    preCond             & = (tmp == iWt[pi][ti] \lor tmp == 0)                \\
                        & \land ((T[ti] \land  tmp == iWt[pi][ti])            \\
                        & \lor (!T[ti] \land tmp == 0))                       \\
    cumulativeIncidentW & = tmp                                               \\
    emptyOutTi          & = !emptyOutTi\}                                     \\
    else\{              &                                                     \\
    preCond             & = preCond \land  (tmp == iWt[pi][ti] \lor tmp == 0) \\
                        & \land ((T[ti]\land tmp == iWt[pi][ti])              \\
                        & \lor  (!T[ti] \land tmp == 0))                      \\
    cumulativeIncidentW & = cumulativeIncidentW + tmp\}
\end{align}
\endgroup
The sum of incident weights at a transition $t_i$ is stored in $cumulativeIncidentW$ as an expression of the $iWt[pi][ti]$ if there is an arc from the transition $t_i$ to place $p_i$ or it is zero. If the transition is not fired, then there are no weights to be considered.

If the expression is non-empty, the previously constructed $preCond$ are \textbf{anded} with the newly constructed expression and the cumulative incident weights are updated in the same manner.

The post condition is constructed if the weight $Wt[pi][ti]!=0$.
\begin{align*}
    postCond          & = ((tmp == Wt[pi][ti]) \lor (tmp == 0) \\
                      & \land((T[ti] \land  tmp == Wt[pi][ti]) \\
                      & \lor (!T[ti]\land  tmp == 0))          \\
    cumulativeWChange & = tmp                                  \\
    emptyChangePi     & = !emptyChangePi
\end{align*}
Based on the above expressions, we construct the transition function $TF$
\begin{verbatim}
    if (emptyTF){
            if (!emptyOutTi){
                TF = preCond & (Px[pi] + cumulativeIncidentW >= 0) 
                emptyTF = false}
            if (!emptyChangePi){                    
                if (!emptyOutTi)
                    TF = TF & postCond 
                    &(Py[pi] == Px[pi] + cumulativeWChange)
                else
                    TF = postCond &(Py[pi] == Px[pi]
                     + cumulativeWChange)
                emptyTF = false
                }
        }
        else{
            if (!emptyOutTi)
                TF = TF &preCond 
                &((Px[pi] + cumulativeIncidentW) >= 0)
            if (!emptyChangePi)
                TF = TF &postCond
                 &(Py[pi] == (Px[pi] + cumulativeWChange))
        }
\end{verbatim}
The transition function is a conjunction of the preconditions, postconditions and the change in the markings. The marking at any instance in time gives the live window characterising the net at that instance.

\subsection{Pre-processing Module}\label{lexParseRules}

In this section we describe the pre-processing in the tool UCSChecker, where the system description given as a $\nu$-net in PNML format and the {\Lstar} formula given in a text file are validated. PNML is based on the Extensible Markup Language, or XML for short, which lends itself to interoperability between tools, while ensuring readability. We employ the ANTLR~\cite{ANTLRParrF11} tool for parsing the two inputs.

\subsection*{Recognizing $\nu$-nets}
We list down the actual \textbf{lexer} and \textbf{parser} grammars that were used to pre-process the $\nu$-nets in the following section.

\subsection*{Lexer rules: }
First, we name the following set of rules, such that we can refer to in the parsing phase:

\begin{lstlisting}
lexer grammar PNMLLexer;
\end{lstlisting}

Consider the following rule that recognizes the escape sequences for tab spaces, carriage return and new line respectively and skips over them when they occur in the input file (PNML file).

\begin{lstlisting}
S           :   [ \t\r\n]+               -> skip ;
\end{lstlisting}

The PNML format is a type of markup file which contains tags. We have the following rules to identify the open, close and comment tags; the digits and text:
\begin{lstlisting}[multicols=2]
COMMENT     :   '<!--' .*? '-->'  -> skip;
SPECIAL_OPEN:   '<?'          -> pushMode(INSIDE) ;
OPEN        :   '<'           -> pushMode(INSIDE) ;
DIGIT       :   [0-9]+ ;
TEXT        :   ~[<&]+ ; //16 bit char except < and &
mode INSIDE;
CLOSE       :   '>'          -> popMode ;
SPECIAL_CLOSE:  '?>'         -> popMode ;
SLASH_CLOSE :   '/>'         -> popMode ;
\end{lstlisting}
We have rules to tokenize the keywords and special characters used in the PNML file:
\begin{lstlisting}[multicols=2]
SLASH       :   '/' ;
EQUALS      :   '=' ;
QUOTE       :   '"' ;
SQUOTE      :   '\'' ;
PLACE       :   'place' ;
TRANSITION  :   'transition' ;
ARC         :   'arc';
INITIAL     :   'initialMarking' ;
INSCRIPTION :   'inscription' ;
TEXTTAG     :   'text' ;
USCORE      :   '_' ;
SOURCE      :   'source' ;
TARGET      :   'target' ;
ID          :   'id' ;
STRING      :   '"' ~[<"]* '"'
            |   '\'' ~[<']* '\'';
\end{lstlisting}
We have a special set of rules for the name tag. The name consists of an alphabet and may be succeded by any combination of digits, alphabets and special symbols (hyphen, underscore, dot), as allowed by the Petri net graphical analysis tool that we used~\cite{PNWolfgang}.
\begin{lstlisting}
Name        :   NameStartChar NameChar* ;
NameChar    :   NameStartChar
            |   '-' | '_' | '.' | DIGIT
            |   '\u00B7'
            |   '\u0300'..'\u036F'
            |   '\u203F'..'\u2040';
NameStartChar  :   [:a-zA-Z]
            |   '\u2070'..'\u218F'
            |   '\u2C00'..'\u2FEF'
            |   '\u3001'..'\uD7FF'
            |   '\uF900'..'\uFDCF'
            |   '\uFDF0'..'\uFFFD';
\end{lstlisting}

\subsection*{Parser rules: }

Now that we have the lexer rules that can identify and tokenize the input, in this section, we write the the parser rules which will be used by ANTLR to construct the parse tree from the valid input and throw errors if any.

First, we set the specific vocabulary of the parser as \textbf{PNMLLexer} (the set of lexer rules we defined above):

\begin{lstlisting}
parser grammar PNMLParser;
options {
    tokenVocab = PNMLLexer;
}
\end{lstlisting}

A valid PNML file contains a header followed by valid elements:

\begin{lstlisting}
doc: header element;
\end{lstlisting}
An element may be either an open tag followed by a close tag or an empty tag, each containing its name and possibly containing a set of attributes. There may be open tags for exactly one of the predefined keywords such as place, transition etc. Notice that it is not possible to define just a close tag using these set of rules. If such an input exists, our tool parses the PNML file and invalidates it by throwing a suitable error.
\begin{lstlisting}
header: '<?' Name attribute* '?>';
element:
       '<' Name attribute* '>' (
        place
        | transition
        | arc
        | element
        | textTagDigit
        | textTag
        | TEXT
        )* '<' '/' Name '>'
        | '<' Name attribute* '/>';
\end{lstlisting}

The place tag contains a mandatory identifier attribute and may contain details to represent the initial marking or special text for parsing.
\begin{lstlisting}
place: '<' 'place' 'id' '=' STRING '>' (initial | element | TEXT)* '<' '/' 'place' '>';
\end{lstlisting}

The rule for parsing the initial marking:
\begin{lstlisting}
initial:
    '<' INITIAL '>' (textTagDigit | textTag | element | TEXT)* '<' '/' INITIAL '>';
\end{lstlisting}

The rule for parsing transitions, with a mandatory identifier.
\begin{lstlisting}
transition:
    '<' 'transition' 'id' '=' STRING '>' (element | TEXT)* '<' '/' 'transition' '>';
\end{lstlisting}
The rule to recognize arcs from a place/transition to transition/place along with their identifiers.
\begin{lstlisting}
arc:
    '<' 'arc' 'id' '=' STRING source target '>' (
        inscription
        | element
        | TEXT
    )* '<' '/' 'arc' '>'
    | '<' 'arc' 'id' '=' STRING source target '/>';
source: 'source' '=' STRING;
target: 'target' '=' STRING;
\end{lstlisting}
Some additional rules to parse the text and specify the type of net for validation:
\begin{lstlisting}
inscription:
    '<' INSCRIPTION '>' (textTagDigit | textTag | element | TEXT)* '<' '/' INSCRIPTION '>';
textTagDigit: '<' TEXTTAG '>' DIGIT '<' '/' TEXTTAG '>';
textTag: '<' TEXTTAG '>' TEXT '<' '/' TEXTTAG '>';
attribute: ('id' | Name) '=' STRING;
\end{lstlisting}

Using these rules above, one can recognize the $\nu$-nets using ANTLR.

\subsection*{Recognizing {\Lstar} properties}
A valid {\Lstar} formula has the server formula $sformula$ which contains the client formula $cformula$ as a sub formula.
\begin{lstlisting}
grammar mlogic;
input :  sformula EOF;
sformula
    : sorop | sandop | simpliesop | suntilop | snotop
        | snextop | sdiamondop | sboxop ;
cformula
    : (corop | candop | cimpliesop | cuntilop | cnotop
        | cnextop | cdiamondop | cboxop );
\end{lstlisting}
We define the server temporal modalities:
\begin{lstlisting}
sorop
    : (exists | forall )(place_x | snotop )
       OR
      ssubformula;
sandop
    : (exists | forall )(place_x | snotop )
       AND
      ssubformula;
simpliesop
    :  (exists | forall )(place_x | snotop )
       IMPLIES
      ssubformula;
suntilop
      : (exists | forall ) (place_x | snotop | snextop | sdiamondop | sboxop)
       SUNTIL
      ssubformula;

snotop
    : NOT (exists | forall ) ssubformula;
snextop
    : SNEXT (exists | forall ) ssubformula;
sdiamondop
    : SDIAMOND (exists | forall ) ssubformula;
sboxop
    : SBOX (exists | forall )  ssubformula;

ssubformula : LPAREN (cformula| place_x | snotop | snextop | sdiamondop
                      | sboxop | sandop | sorop | simpliesop | simpliesop) RPAREN;
\end{lstlisting}
Consequently, we define the client temporal modalities:
\begin{lstlisting}
corop
    : (place_x | cnotop | cnextop | cdiamondop | cboxop)
        OR
        csubformula;
candop
    : (place_x | cnotop | cnextop | cdiamondop | cboxop)
        AND
        csubformula;
cimpliesop
    : (place_x | cnotop | cnextop | cdiamondop | cboxop)
        IMPLIES
        csubformula;
cuntilop
    : (place_x | cnotop | cnextop | cdiamondop | cboxop)
        CUNTIL
        csubformula;
cnotop
    : NOT csubformula;
cnextop
    : CNEXT csubformula;
cdiamondop
    : CDIAMOND csubformula;
cboxop
    : CBOX csubformula;

csubformula: LPAREN (place_x | cnotop | cnextop | cdiamondop
                     | cboxop | candop | corop | cimpliesop | cuntilop) RPAREN;
\end{lstlisting}
We provide the rules governing quantifiers:
\begin{lstlisting}
exists : EXISTS VAR;
forall : FORALL VAR;
place_x : PLACE DIGIT LPAREN VAR RPAREN;
\end{lstlisting}
We recognize the valid tokens for the various temporal modalities and operators:
\begin{lstlisting}
EXISTS: 'E';
FORALL: 'V';
PLACE : 'p';
NOT : '~';
SNEXT : 'X_s';
SDIAMOND : 'F_s';
SBOX : 'G_s';
CNEXT : 'X_c';
CDIAMOND : 'F_c';
CBOX : 'G_c';
OR : '|';
AND : '&';
IMPLIES : '%';
SUNTIL : 'U_s';
CUNTIL : 'U_c';
\end{lstlisting}
We recognize the valid symbols such as paranthesis and identifiers and spaces:
\begin{lstlisting}
LPAREN : '(';
RPAREN : ')';
NAME : '\'' ~[']+ '\'';
VAR: [a-z];
DIGIT : '-'?[0-9]+;
ENDLINE : ('\r\n'|'\n'|'\r')+ -> skip;
WHITESPACE : [\t ]+ -> skip;
\end{lstlisting}

Instead of providing separate parser and lexer rules, we provide the grammar for recognizing {\Lstar} in one shot. Using the above rules, we validate the {\Lstar} specifications in the accompanying tool.

\end{document}